\documentclass{article}
\usepackage{arxiv}
\usepackage[utf8]{inputenc}
\usepackage[T1]{fontenc}
\usepackage{amsmath}
\usepackage{newtxtext,newtxmath}
\usepackage{algorithm}
\usepackage{algorithmic}

\usepackage{makecell}
\usepackage{gensymb}
\usepackage{xcolor}
\usepackage[colorlinks=true, allcolors=blue]{hyperref}
\usepackage{xurl}
\usepackage{booktabs}
\usepackage{microtype}
\usepackage{graphicx}
\usepackage[numbers,sort&compress]{natbib}
\usepackage[font=small,labelfont=bf]{caption}

\graphicspath{{Figures/}{../Figures/}}
\newcommand*{\doi}[1]{doi:\,\href{https://doi.org/#1}{\nolinkurl{#1}}}

\title{A spinal-inspired spiking neural network for continuous\\ finger-joint kinematics estimation\\ from high-density surface EMG}

\newcommand{\orcidlink}[1]{\href{https://orcid.org/#1}{\includegraphics[width=9pt]{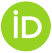}}}
\author{%
  Cristian David Guerrero-Mendez$^{1,2}$\,\orcidlink{0000-0003-1019-9423}
  \qquad
  Leonardo Abdala Elias$^{1,2,*}$\,\orcidlink{0000-0003-4488-3063}\\[8pt]
  \normalfont\small
  $^{1}$Department of Electronics and Biomedical Engineering, School of Electrical and Computer Engineering,\\
  \normalfont\small University of Campinas (UNICAMP), Campinas, SP 13083-852, Brazil\\[3pt]
  \normalfont\small
  $^{2}$Neural Engineering Research Laboratory, Center for Biomedical Engineering,\\
  \normalfont\small University of Campinas (UNICAMP), Campinas, SP 13083-881, Brazil\\[3pt]
  \normalfont\small
  $^{*}$Corresponding author: \href{mailto:leoelias@unicamp.br}{leoelias@unicamp.br}
}
\date{}

\renewcommand{\shorttitle}{Spinal-inspired SNN for finger-joint kinematics estimation from HD sEMG}

\hypersetup{
pdftitle={A spinal-inspired spiking neural network for continuous finger-joint kinematics estimation from high-density surface EMG},
pdfauthor={Cristian David Guerrero-Mendez, Leonardo Abdala Elias},
pdfkeywords={electromyography, fingers, kinematics, neural networks, spiking neural network, simultaneous and proportional prediction},
}

\begin{document}
\maketitle

\begin{abstract}
\emph{Objective:} Decoding simultaneous and proportional finger kinematics from high-density surface electromyography (HD sEMG) remains challenging. Although data-driven methods have shown promising performance for finger movement decoding, their limited physiological interpretability remains a relevant limitation, particularly because physiologically grounded elements are often absent from their architectures. This study proposes and evaluates a spiking neural network (SNN) that incorporates spinal-inspired structures and mechanisms to predict multi-joint finger kinematics from HD sEMG signals. \emph{Approach:} The spinal-inspired SNN includes mechanisms such as a spike-encoding computational model with input-noise variability, a phenomenological representation of Henneman’s size principle, and excitation/inhibition (E/I) balance. HD sEMG and kinematic data were collected from 23 healthy participants performing grasping tasks under different elbow and wrist postures. Performance was assessed using six prediction metrics and compared against six baseline machine learning and deep learning methods. Non-parametric statistical analyses were performed to evaluate differences in the physiological explorations and performance comparisons. \emph{Main results:} The proposed method showed a large and significant ($p<$ 0.05, rank-biserial correlation $r_{rb}>$ 0.7) improvement over all baseline methods across all accuracy metrics. Performance improved significantly with increasing spike encoder firing rates up to approximately 83 Hz (encoder amplitude $\theta$ = 250, beyond which performance reached a plateau), and more regular input spike trains compared with a Poisson-like encoder. In contrast, the E/I balance did not significantly affect the decoding performance of the algorithm. The firing rate and inter-spike interval statistics of the SNN differed significantly across motor tasks. \emph{Significance:} These results demonstrate that incorporating neurophysiological principles into SNNs improves prediction performance while providing additional physiological interpretability for simultaneous and proportional finger-kinematics decoding. This approach represents a promising step toward physiology-informed neural networks, with potential applications in motor rehabilitation.
\end{abstract}


\section{Introduction}
The human hand plays a central role in performing daily tasks, supporting not only grasping and object manipulation but also haptic exploration \cite{Lederman1987}, skilled tool use \cite{Maravita2004}, expressive actions through manual gestures \cite{Clough2020}, and social communication through gesture and touch \cite{Kirsch2018}. These functions rely on the precise coordination of multiple fingers, while individuated finger control enables highly skilled actions such as writing \cite{Bisio2016}, drawing or painting \cite{Zarandi2023}, and musical performance \cite{Kimoto2019}. Natural hand movements arise from the interaction between independent finger control and synergistic neural commands that coordinate the fingers into smooth, refined motor patterns \cite{Schieber2004}. Therefore, accurately decoding hand and finger kinematics from neuromuscular signals is highly relevant for developing intuitive human-machine interfaces and neurorehabilitation strategies.

Data-driven methods have been extensively applied in motor rehabilitation, using muscle or brain signals as interfaces \cite{Benzy2020, Simpetru2023, Madduri2026}. Despite their potential impact, several challenges remain in both neurorehabilitation interventions and data-driven approaches. On the one hand, regarding interventions, it is still unclear: (i) how to interpret the underlying neuromuscular or neurophysiological mechanisms involved during the intervention \cite{Buma2013}; (ii) how to evaluate their effectiveness using neural-based metrics to support the development of personalized strategies; and (iii) which neural-based measures could serve as reliable references for guiding interventions \cite{Garro2021}. On the other hand, data-driven methods face additional challenges, including: (i) a strong dependence on large amounts of data \cite{Halmich2025}, which are often difficult to obtain in clinical settings; (ii) performance variability associated with signal processing choices, feature extraction, and model architecture \cite{Simpetru2023}; and (iii) the limited incorporation of physiologically grounded elements into their architectures, which restricts their interpretability from a neuromuscular perspective. These challenges highlight the need to develop methodological alternatives based on physiologically inspired approaches that can improve the simultaneous and proportional decoding of kinematic variables during movement execution.

A promising solution involves designing methods with more physiologically accurate representations within data-driven frameworks to address challenges associated with interventions and data-driven approaches. Recent studies have proposed physics-informed deep learning approaches to predict muscle force and joint kinematics \cite{Ma2024, Zhang2023, Shi2024, Taneja2024}, demonstrating that incorporating physical laws into training can match or exceed the performance of traditional machine learning methods. Beyond traditional deep learning, spiking neural networks (SNNs) offer a more biologically plausible framework that represents neural information as discrete spike events while incorporating membrane potential dynamics, subthreshold synaptic integration, and the temporal accumulation of inputs \cite{Yamazaki2022}. However, most SNN applications in sEMG signal processing have focused on gesture classification tasks \cite{Sun2023, Vitale2022}, often relying on relatively simple architectures with limited biological realism and primarily targeting neuromorphic implementations \cite{Tanzarella2023, Ceolini2020, Donati2019, Vitale2022}. Although these approaches were evaluated in upper-limb motor tasks involving various hand movements, their application to gesture discrimination limits knowledge of the ability to predict natural, continuous movements. More recently, SNNs have been applied to the continuous regression of finger kinematics from sEMG, using spiking transformers for low-latency prosthetic control \cite{Leroux2023}, heterogeneous populations of leaky integrate-and-fire (LIF) neurons with variable thresholds and time constants to encode sEMG \cite{Baracat2025}, and benchmarks of temporal models across subjects \cite{Manneschi2026}. In a related direction, shallow SNNs have decoded individual finger forces from motor unit spike trains and spike-encoded intramuscular EMG recorded with high-density microelectrode arrays \cite{Baracat2026}. However, these approaches target computational efficiency, encoding robustness, or cross-subject generalization, without organizing the network according to the principles of spinal motor control.

In parallel, other studies have explored physiologically grounded computational models as alternatives to purely data-driven approaches. For example, Maksymenko \textit{et al.} proposed the concept of a myoelectric digital twin \cite{Maksymenko2023}, a physiologically realistic and computationally efficient simulation framework that generates synthetic data for training deep learning algorithms. In a related direction, Caillet et al. \cite{Caillet2023} implemented a motor neuron pool model calibrated using high-density surface electromyography (HD sEMG) signals, which described force-generation dynamics through a Hill-type muscle model driven by the identified motor neuron activity. The latter approach enabled the extraction of subject- and muscle-specific neuromechanical information, providing a more physiological description of motor unit activation and improving muscle force prediction. Similarly, motor neuron pool models with explicit recruitment \cite{Fuglevand1993} and biophysically detailed simulators of spinal cord circuits \cite{Cisi2008, Elias2012} describe how descending commands are transformed into motor unit activity and muscle force. Still, they operate as forward-simulation models rather than as trainable decoders. These studies support the development of computational methodologies that move beyond conventional data-driven approaches by integrating biophysical and neuromechanical models, thereby enabling a more comprehensive representation of complex biological processes. However, to the best of our knowledge, the use of physiologically inspired models for the simultaneous and proportional decoding of continuous hand and finger kinematics remains less explored.

In this study, we introduce a spinal-inspired SNN that incorporates neurophysiological principles of spinal motor control and motor neuron pool organization. The method integrates a spike-based encoding of the sEMG input, a phenomenological representation of motor neuron recruitment inspired by Henneman’s size principle, excitation/inhibition (E/I) balance, and stochastic input variability. We evaluated how these physiologically inspired components influence the prediction of 15 hand joint angles from HD sEMG signals recorded from three motor tasks and multiple upper-limb postures. To the best of our knowledge, this study is among the first to incorporate principles of spinal motor control into a trainable SNN to enable simultaneous and proportional decoding of continuous finger joint movements.

\section{Methods}
\subsection{Experimental dataset}
Twenty-three healthy volunteers participated in the study. Of these, twelve identified as male and eleven as female, with twenty-one being right-handed and two left-handed (both male). All recordings were performed on the right hand. Participants were recruited across different age ranges, from 20 to 70 years, organized in 10-year intervals. Each age group included at least two male and two female participants. Overall, the sample had a mean age of 42.30$\pm$14.85 years, body mass of 75.04$\pm$14.23 kg, height of 167.91$\pm$9.28 cm, and forearm circumference of 26.29$\pm$2.70 cm. All participants were fully informed about the experimental procedures and voluntarily signed an informed consent form prior to their inclusion in the study. The experimental protocol was approved by the Research Ethics Committee of the University of Campinas (UNICAMP), under protocol number CAAE: 80299124.2.1001.5404, and was conducted in accordance with the Declaration of Helsinki for research involving human subjects. Metadata describing the database is available in \cite{Guerrero2026}.

\subsubsection{Signal acquisition}
Simultaneous recordings of HD sEMG signals were obtained from proximal forearm muscles --- Flexor Digitorum Superficialis (FDS) and Extensor Digitorum Communis (EDC) --- together with hand kinematics estimated using a multicamera system \cite{Cakici2022} (see Figure \ref{fig:Protocol}A). Specifically, muscle activity was acquired using two electrode grids of 64 channels each (8 $\times$ 8 configuration, 10 mm inter-electrode distance; GR10MM0808, OT Bioelettronica, Italy), operating in monopolar mode with a sampling frequency of 2052 Hz, using the wearable HD sEMG acquisition platform described in \cite{Molinari2026}, based on two 64-channel RHD2164 amplifiers (Intan Technologies, USA). Kinematic recordings were performed using five high-speed cameras (DFK-37BUX287, The Imaging Source$^{TM}$) equipped with 5 MP lenses (TCL 0814, The Imaging Source$^{TM}$), arranged around a cubic structure of 0.34 m$^3$. The cameras captured hand motion at 100 Hz (FPS) with a resolution of 720 × 540 pixels, using IC Capture software (The Imaging Source$^{TM}$). Simultaneously, to regulate the frequency of movement execution, a virtual interface was presented to the participants on a monitor located approximately 1 m in front of them. This interface displayed a virtual hand indicating both the type of movement and the frequency at which the task should be performed. The interface was developed using the Unity Engine (Unity Technologies, USA) and presented movements at a frequency of 0.5 Hz.

Synchronization between the acquisition systems and the virtual interface was ensured through a command station that sent trigger pulses to the cameras and simultaneously transmitted a synchronization signal via Wi-Fi over TCP/IP to the sEMG system and Unity's interface.

\subsubsection{Procedures}
The acquisition procedure began with an initial assessment of the participant's physical characteristics, history of injuries, and physical activity. Subsequently, each participant completed tests of laterality and manual dexterity using the Edinburgh Handedness Inventory \cite{Oldfield1971} and the Nine Hole Peg Test \cite{Mathiowetz1985}. However, the results of these tests are out of the scope of the present report.

\begin{figure}
    \centering
    \includegraphics[width=0.65\linewidth]{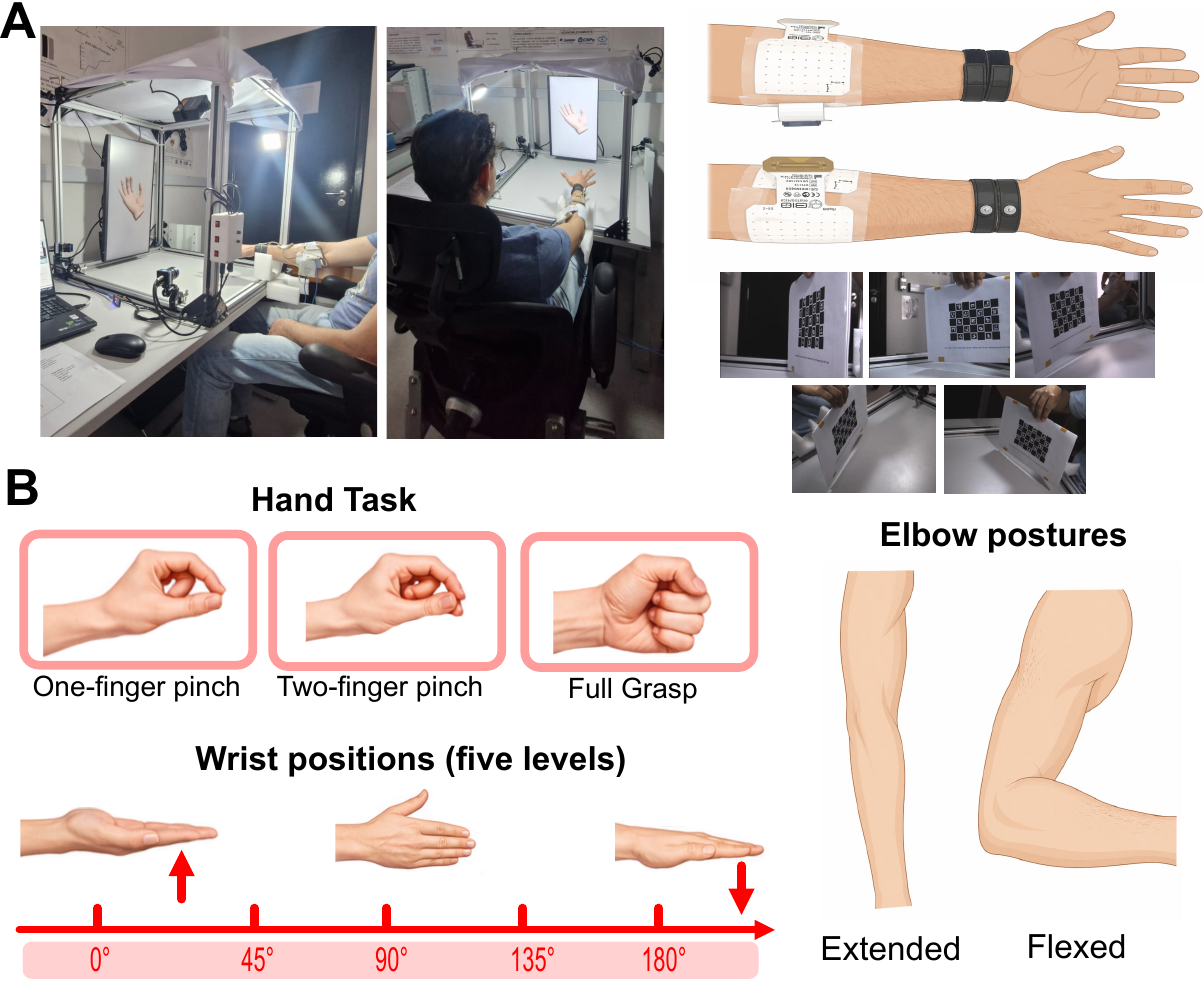}
    \caption{A) Experimental setup, grid placement, and camera calibration using a ChArUco board. B) Experimental protocol, showing the hand tasks, the elbow postures (flexion--extension) and wrist positions (supination--pronation; 0$^{\circ}$: supination, 180$^{\circ}$: pronation).}
    \label{fig:Protocol}
\end{figure}

Prior to data acquisition, participant preparation included cleaning the forearm skin with light abrasion using an exfoliating paste, followed by application of 70\% alcohol. The two HD sEMG electrode grids were placed over the FDS and EDC muscles following a standardized protocol \cite{Zipp1982}, and secured with adhesive bandages to ensure steady contact throughout the experimental session. Reference electrodes were positioned on the wrist using a water-moistened conductive band. Additionally, prior to kinematic recordings, the camera system was spatially calibrated for 60 s using a ChArUco board (see Figure \ref{fig:Protocol}A).

Participants performed finger movement tasks in a seated position (see Figure \ref{fig:Protocol}A). The tasks included a one-finger pinch (1FP), a two-finger pinch (2FP), and a full-hand grasp (FG). The tasks were combined with five wrist postures, corresponding to the combined rotation of the forearm and wrist (supination--pronation) from 0$^{\circ}$ (supination) to 180$^{\circ}$ (pronation) in 45$^{\circ}$ increments, and two elbow postures (flexed and extended), resulting in a total of 30 experimental conditions (tasks $\times$ wrist postures $\times$ elbow postures) (see Figure \ref{fig:Protocol}B). For conditions involving an extended elbow, slight variations in the distance between the participant and the monitor were required. Due to the cameras' limited field of view, participants were positioned slightly farther back under these conditions to ensure the hand remained within the frame, thereby avoiding the need for system recalibration. For each participant, the order of task execution was fully randomized, and each condition was performed for 30 s. Note that this corresponds to a total of 15 repetitions per task, given that the movement frequency was set to 0.5 Hz in the interface. Participants were instructed to follow as closely as possible the sinusoidal finger movements displayed by the virtual hand in Unity. All tasks began with the fingers in full extension, and a minimum rest period of 30 s was provided between conditions. On average, each acquisition session lasted approximately 100 minutes per participant.

Additionally, participants were instructed to perform the movements naturally, using their full range of motion (ROM), while avoiding hyperextension, forced flexion, or excessive force; that is, only light contact between the fingertips was required. They were also instructed to initiate each movement with the thumb fully extended, followed by complete opposition and retraction. The experimenter supervised these instructions individually before each session to ensure proper task execution. In cases of incorrect performance, loss of hand visibility within the camera's field of view, or any other issues that compromised data quality, the tasks were repeated.

\subsection{Data pre-processing}
\begin{itemize}
    \item \textit{HD sEMG signals}: HD sEMG signals were preprocessed over the entire recording using a causal notch filter centered at 60 Hz to attenuate power-line interference, followed by a fourth-order causal Butterworth band-pass filter with cutoff frequencies of 10 and 500 Hz. Because both filters were causal, each filtered sample depended only on the current and past samples, which prevented information leakage from the validation and test sets into the training set. Muscle activation envelopes were subsequently obtained by computing the root-mean-square (RMS) of each HD sEMG channel using a causal normalized rectangular sliding window of 300 ms. The initial 2 s of each recording were then discarded to exclude the transient at task onset, during which participants synchronized their movements with the virtual hand, as well as the start-up transient of the causal filters; this interval corresponds to one complete movement cycle at the 0.5 Hz movement frequency. The remaining data were subsequently divided chronologically into training, validation, and test sets: the last 3 s formed the test set, the preceding 3 s the validation set, and the remaining samples (approximately 22 s) the training set. Min–max scaling was applied to the RMS signals, with the scaling parameters estimated exclusively from the training set and then applied to the validation and test sets.

    \item \textit{Joint Kinematics}: For the reconstruction of three-dimensional (3D) finger kinematics from the recordings obtained with the five cameras, the MediaPipe Hand framework \cite{MediaPipe} was first used to independently process the recordings from each camera, detecting and tracking 21 virtual hand skeleton landmarks in 2D coordinates. Subsequently, Anipose \cite{Karashchuk2021} was used to reconstruct 3D trajectories from 2D trajectories. The reconstruction initially relied on calibration recordings acquired prior to the experiments to estimate the intrinsic and extrinsic parameters of each camera. Then, the triangulation was performed to obtain the 3D coordinates ($x$, $y$, $z$) of each marker at every time point. Triangulation retained MediaPipe detections with a confidence score $\geq$ 0.6 and applied the spatiotemporal optimization of Anipose with segment-length constraints. The quality of the 3D reconstruction was assessed by the reprojection error of the camera calibration and of the hand landmarks, the latter reflecting the agreement of the 2D detections across cameras.

    Joint angles were computed for the carpometacarpal (CMC), metacarpophalangeal (MCP), thumb interphalangeal (IP), and proximal (PIP) and distal (DIP) interphalangeal joints, yielding three angles per finger (see Figure \ref{fig:Model}). The computation of angles at the MCP and thumb carpometacarpal joints employed an orthogonal projection strategy, as these joints are not constrained to a fixed plane due to abduction/adduction, which can affect flexion/extension estimation. To this end, two vectors were defined using three markers: proximal, central (target joint), and distal. These vectors were then projected onto a plane perpendicular to a direction vector $\vec{d}$, defined toward the MCP of the nearest adjacent finger. For the thumb, $\vec{d}$ was estimated using virtual markers located at the wrist. The joint angle was then computed between the projected vectors using the $atan2(\cdot)$ function. For interphalangeal joints, where motion occurred approximately within a single plane, no projection was applied. The angle was computed directly between two vectors defined by three markers, also using the $atan2(\cdot)$ function. For all joints, 0$^{\circ}$ corresponded to full extension (aligned segments), and angles increased with flexion. A similar biomechanical model was recently applied with a marker-based optical motion-capture system \cite{AvilsCarrillo2026}. 

    Following angle estimation, the kinematic data were resampled by linear interpolation to match the sEMG sampling rate, and the same initial 2 s trimming and division into training, validation, and test sets as for the HD sEMG signals were applied.  Finally, the angle signals were standardized using z-score normalization, with the mean and standard deviation computed exclusively from the training data and subsequently applied to the validation and testing sets.
\end{itemize}

\subsection{The spinal-inspired spiking neural network}
\begin{figure}
    \centering
    \includegraphics[width=0.95\linewidth]{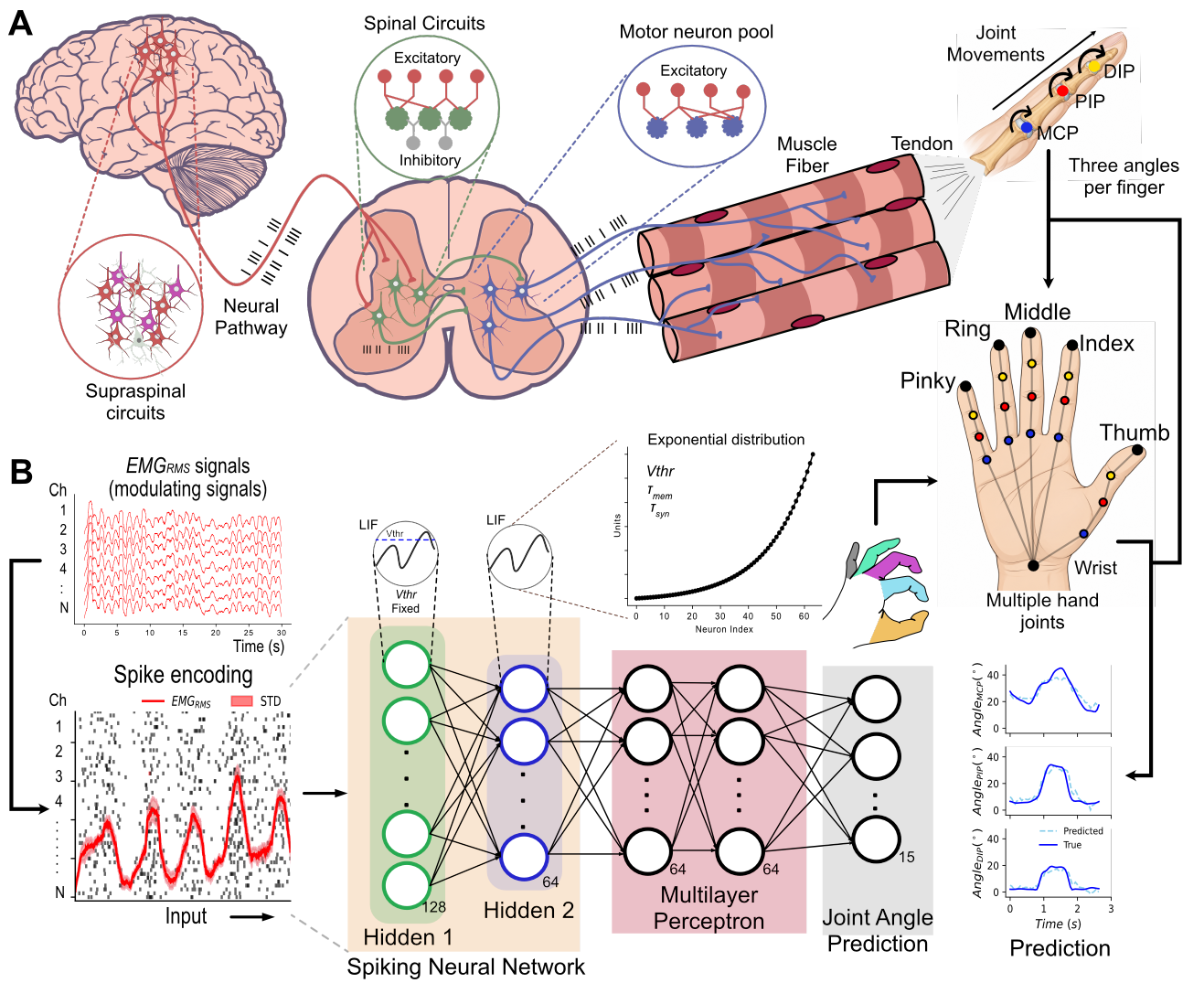}
    \caption{Schematic overview of the spinal-inspired SNN and its relationship with the physiological process during motor control. A) A biological representation of the neural pathways that extend from the brain, through the spinal circuits, until reaching the muscle fibers, thereby generating force and joint movements. The illustrations of the human brain, spinal cord, muscle fibers, and neurons were obtained and adapted from Scidraw.io (doi.org/10.5281/zenodo.3926243, doi.org/10.5281/zenodo.4966122, doi.org/10.5281/zenodo.7668290, doi.org/10.5281/zenodo.3925929, doi.org/10.5281/zenodo.3925905). B) Computational model inspired by each stage of the process, where spike encoding converts the sEMG envelope into spike trains, loosely analogous to the descending drive to the spinal cord, the SNN models a simplified version of spinal circuits, and the multilayer perceptron (MLP) maps the SNN activity to the joint angles. In this case, the outputs correspond to fifteen hand joint angles. MCP: Metacarpophalangeal joint; PIP: Proximal interphalangeal; DIP: Distal interphalangeal.}
    \label{fig:Model}
\end{figure}

An overview of the implementation of the spinal-inspired SNN ($SNN_{Physio}$) and its relationship with biological processes occurring in the central nervous system is presented in Figure \ref{fig:Model}. The method was implemented in Python using PyTorch \cite{PytorchSoft} (v2.5.1).

\subsubsection{Spike Encoding:}
Various encoding schemes have been proposed for SNNs, particularly in sEMG-based gesture recognition and neuromorphic implementations. Among these schemes, temporal and rate-based encoders are commonly used \cite{Sun2023, Vitale2022, Ceolini2020, Petro2020}. However, although these approaches are computationally effective, they do not adequately capture the biological properties of neuronal firing patterns. Therefore, in this work, we propose an encoding model based on a non-homogeneous renewal point process (Poisson or Gamma) \cite{Perkel1967, Gabbiani2017}. This model emulates a neuron that integrates incoming input over time through an internal state variable. Spike generation occurs when this variable exceeds a stochastic threshold with unit mean, and the variability is determined by the process order \cite{Barbieri2001}. Following each spike event, both the accumulated state and the threshold are reset, ensuring independence between successive events.

As input to the encoder, the normalized RMS signal from each HD sEMG channel is used. This signal is scaled by a constant gain ($\theta$), which converts the normalized RMS amplitude into an instantaneous firing rate. Processing is performed in a temporal manner for each channel, evaluating the integration dynamics of the proposed neuronal model at each time step (one sEMG sample, $1/2052$ s $\approx$ 0.49 ms). Additionally, the model incorporates an order parameter that controls the regularity of the spike train: when the order is 1, the behavior corresponds to a Poisson process (highly irregular), whereas higher order values yield more regular discharge, resulting in a Gamma point process. Because the threshold has a unit mean regardless of the order, this parameter modifies spike-train regularity without altering the mean firing rate. In this way, different input variability (noise) in event generation can be effectively adjusted.

As a result, the encoder produces spike trains organized in a matrix of dimensions $C \times N$, where $C$ represents the number of sEMG channels and $N$ denotes the number of temporal samples. This representation can be interpreted as a population of input spike trains over time, whose firing rate is directly influenced by the amplitude of the input signal (see Figure \ref{fig:Model}B). The properties and pseudo-code of the spike encoder are provided in Section 2 of the \textit{Supplementary Material}.

\subsubsection{Neuron model:}
Each neuron in the SNN was modeled using a LIF model \cite{Brunel2007}. Here, an extended LIF formulation was adopted, accounting for decay mechanisms in both the membrane potential and the input current in the voltage dynamics (Equation \ref{Eq:Voltage}), thereby incorporating leakage effects into the subthreshold response.

\begin{equation} \label{Eq:Voltage}
    \tau_{mem}\frac{d V(t)}{dt} = -V(t) + R I(t),
\end{equation} where $V(t)$ denotes the membrane potential, $I(t)$ corresponds to the synaptic current resulting from spike integration, $\tau_{mem}$ represents the membrane time constant, and $R$ denotes the associated membrane resistance.

To increase the biological realism of the neuronal model, the synaptic current dynamics were represented using a first-order approximation that captures its exponential decay (Equation \ref{Eq:Current}).

\begin{equation}\label{Eq:Current}
    \tau_{syn}\frac{dI(t)}{dt} = -I(t) + \tau_{syn} W S(t),
\end{equation} where $W$ corresponds to the synaptic weight matrix, $S(t)$ represents the input spike trains, and $\tau_{syn}$ defines the time constant associated with synaptic decay. Unlike the classical LIF model, which only accounts for membrane potential dissipation, this formulation introduces an additional dynamic by explicitly modeling the temporal evolution of the synaptic current, which is governed by its own time constant.

For computational implementation, the neuron dynamics were expressed in discrete time and integrated with the feedforward connections between layers. At each time step, the synaptic current and membrane potential of neuron ($i$) in layer ($l$) were updated (Equations \ref{Eq:VoltageDes} and \ref{Eq:CurrentDes}).

\begin{equation} \label{Eq:VoltageDes}
V_{i}^{(l)}(t) = \tau V_{i}^{(l)}(t - 1) + (1-\tau) R I_{i}^{(l)}(t),
\end{equation}

\begin{equation} \label{Eq:CurrentDes}
I_{i}^{(l)}(t) = \mu I_{i}^{(l)}(t - 1) + \sum_{j} w_{i,j} S_{j}^{(l-1)}(t) + b_i^{(l)},
\end{equation} where $\tau = e^{-\frac{1}{\tau_{mem}}}$ and $\mu = e^{-\frac{1}{\tau_{syn}}}$ are dimensionless exponential decay factors controlling the decay of the membrane potential and synaptic current, respectively. Equation~\ref{Eq:VoltageDes} corresponds to the exact discretization of Equation~\ref{Eq:Voltage}. In the implementation, the input coefficient $(1-\tau)R$ was set to 1 without loss of generality, since the learnable weights compensate any scaling of the input current; thus, $V$ and $V_{thr}$ are expressed in arbitrary units. In the present implementation, $\Delta t$ = 1 discrete time step, with $\tau_{mem}$=10 and $\tau_{syn}$=5 time steps ($\approx$4.9 and $\approx$2.4 ms, respectively) for the first hidden layer; the corresponding values for the second layer follow the exponential distribution described in Section~\ref{sec:PhysioComponents}. In this notation, $l$ denotes the current layer, while $l - 1$ refers to the preceding layer. The index $j$ corresponds to the $j$-th neuron in layer $l - 1$, and $S_{j}^{l-1}$ denotes the associated input spikes. Finally, $w_{i,j}$ describes the synaptic connection from neuron $j$ in the previous layer to neuron $i$ in layer $l$, and $b_i^{(l)}$ denotes a learned neuron-specific bias input that adjusts the baseline excitability of the neurons.

Finally, when the membrane potential $V_{i}^{(l)}(t)$ reaches the threshold value ($V_{thr}$), the corresponding spike event is registered, as defined in Equations \ref{Eq:Spikes} and \ref{Eq:Func}.

\begin{equation} \label{Eq:Spikes}
    S_{i}^{(l)}(t) = f  \left ( V_{i}^{(l)}(t) - V_{thr} \right ),
\end{equation}

\begin{equation} \label{Eq:Func}
    f(x) = \left\{\begin{matrix}
            0, x < 0  \\
            1, x \geq 0 \\
            \end{matrix}\right.
\end{equation}

After spike generation, the membrane potential is updated using a reset mechanism defined in Equation \ref{Eq:Reset}. This process is controlled by a penalty factor $P$, which regulates the magnitude of the neuronal state reset. Depending on this parameter, the membrane potential can either return to zero or take negative values that emulate hyperpolarization. In the present implementation, $P = 1$ is used, resulting in a full reset of the membrane potential to zero after each event. It is worth noting that this type of LIF neuron has recently been employed in hand gesture classification tasks \cite{Sun2023}. Finally, an analysis of the behavior of the implemented LIF model is presented in Section 1 of the \textit{Supplementary Material}.

\begin{equation} \label{Eq:Reset}
    V_{i}^{(l)}(t) = V_{i}^{(l)}(t) (1 - P S_{i}^{(l)}(t))
\end{equation}

\subsubsection{Architecture:}
The SNN architecture (Figure~\ref{fig:Model}B) consists of an input stage based on spike encoding, as previously described, followed by two hidden layers of LIF neurons that emulate spinal circuits, and a final stage comprising a multilayer perceptron (MLP) that generates a continuous representation for finger movement prediction.

The input layer consists of 128 neurons modulated by an amplitude parameter $\theta$ (spike encoding). The two hidden layers, responsible for extracting spatiotemporal patterns, contain 128 and 64 LIF neurons, respectively. In the output stage, the network summarizes temporal information by using the membrane potential of the last hidden layer averaged over a 100 ms integration window (50 ms step, 50\% overlap) for each neuron. Neuronal states were reset to zero at the start of each window, and the membrane potential was taken after the reset (Equation~\ref{Eq:Reset}). This representation is then fed into a two-layer MLP: the first layer projects the features from the last SNN layer into a 64-dimensional space, followed by a Rectified Linear Unit (ReLU) activation and a dropout layer with a rate of 0.2 to improve generalization. Finally, a second linear layer maps these representations into a 15-dimensional space corresponding to all finger joints, thereby producing continuous outputs for joint-movement decoding. The corresponding kinematic target for each window was taken to be the joint angle at the last sample of that window. 

The number of neurons in the hidden layers was experimentally tuned to balance the model’s ability to represent the input patterns and its computational cost, avoiding an excessively complex architecture.

\subsection{Physiologically inspired components}\label{sec:PhysioComponents}
In addition to the physiological incorporation introduced in the spike-encoding model, other modifications were implemented to analyze behavior under a more biologically inspired representation of spinal circuits. 

\begin{itemize}
    \item \textit{Exponential distribution of neurons}: This adaptation was implemented by modifying the firing threshold, and membrane and synaptic time constants of the neurons in the second layer of the SNN. The thresholds followed an exponential distribution (Figure~\ref{fig:Model}B) between $V_{thr}-2$ and $V_{thr}+2$, where $V_{thr}$ is the baseline threshold tuned as a hyperparameter and corresponds to the midpoint of this range. The offset of $\pm 2$ was selected empirically to keep all thresholds positive within the explored values of $V_{thr}$. The same exponential interpolation scheme was also applied to the membrane and synaptic time constants of the second layer ($\tau_{mem}$ and $\tau_{syn}$, ranging from 16 to 6 and from 8 to 3 time steps, respectively; i.e., $\approx$7.8--2.9 ms and $\approx$3.9--1.5 ms), following the same neuron ordering used for the threshold distribution: neurons with the lowest threshold (recruited first) were assigned the longest time constants, while neurons with the highest threshold (recruited last) were assigned the shortest ones. This graded organization is consistent with the inverse relationship between motor neuron size, input resistance, and membrane time constant described within Henneman's size principle \cite{Henneman1965}; thus, this distribution constitutes a phenomenological representation of the size principle, whereas the first layer retained fixed time constants ($\tau_{mem}$ $\approx$ 4.9 ms, $\tau_{syn}$ $\approx$ 2.4 ms). 

    \item \textit{Spike encoder firing rate}: Increasing the amplitude ($\theta$) of the modulating signal in the spike encoder is directly proportional to an increase in the firing rate at the encoder output (see Section 2 of the \textit{Supplementary Material}).

    \item \textit{Variability of input spikes}: The spike encoder implemented in the SNN allows control over the variability of spike trains through the order parameter. Accordingly, different levels of variability in the encoding process were analyzed. This enables indirect control of the noise level at the SNN input: lower orders generate more irregular spike trains (i.e., higher noise), whereas higher orders produce more regular patterns associated with lower noise levels. For instance, when the order is 1, the spike encoder yields a Poisson point process, with a coefficient of variation (CoV) equal to 1, indicating fully irregular firing activity. As the order increases, the CoV decreases following $\left(\frac{1}{\sqrt{n}}\right)$, where $n$ denotes the model order.

    \item \textit{Excitation/inhibition balance}: To implement the E/I balance, the sign of the synaptic weights in the first layer of the SNN was fixed per presynaptic input channel (spikes from the spike encoder for each channel) rather than per individual connection, so that each input projected exclusively excitatory or exclusively inhibitory connections to all postsynaptic neurons, consistent with Dale's principle \cite{Strata1999}. Sign was assigned through a random permutation of the input channels, while connection magnitudes reused the network's default initialization and remained the only trainable component, via the reparameterization $w=\text{sign}\times|w_{raw}|$; gradient descent could therefore adjust magnitude but never invert sign, so the assigned E/I balance was exactly fixed throughout training.
\end{itemize}

\subsection{Hyperparameter tuning and training}
The hyperparameter selection process focused on four key variables: $V_{thr}$, $\theta$, the encoder order, and the E/I balance. These hyperparameters were optimized using Optuna (v4.9.0) \cite{Akiba2019}, an open-source framework for automated hyperparameter optimization, over the following search space: $V_{thr}$ from 3 to 6 (step of 0.1); $\theta$ of 200, 250, 300, 350, and 400; order of 5, 15, and 25; and E/I balance of 65/35, 70/30, 80/20, or without balance (WB). The search was guided by the tree-structured Parzen estimator (TPE) sampler \cite{Bergstra2011} with Optuna's default settings. Each optimization study was run sequentially under a fixed budget of 60 trials, and the objective function was defined as maximizing the variance-weighted coefficient of determination ($R^2$) computed on the validation set. To ensure reproducibility, the seed of both the sampler and the model initialization was fixed across all studies. Each trial followed the training protocol described below.

Tuning was performed separately for each task in five participants, one randomly selected from each age group (20–70 years, in 10-year intervals), with the elbow extended and the wrist at 90$^{\circ}$, the neutral forearm position of the protocol (15 Optuna calibrations). These participants were also included in the final analysis, given that tuning used only the validation set and test sets were never used for model selection. The hyperparameter combinations evaluated in each calibration (Optuna trials) were ranked according to the variance-weighted $R^2$, and the final combination for each task was selected from the best-ranked configurations across participants. Section 4 of the \textit{Supplementary Material} presents the hyperparameter optimization results and the selection procedure.

The SNN was trained end-to-end with backpropagation through time and a surrogate gradient \cite{Neftci2019}, implemented directly in PyTorch. Training was performed using the AdamW optimizer, with a constant learning rate of $\eta = 10^{-3}$ and a decoupled weight decay of $\lambda = 10^{-3}$. The learning rate was kept fixed across all epochs. The training process was carried out for a maximum of 100 epochs, incorporating an early stopping criterion if no improvement in the validation loss was observed for 13 consecutive epochs, where any decrease in the validation loss counted as an improvement. The final trained model corresponds to the weights from the epoch that achieved the lowest validation loss. The cost function used was the L1 loss, corresponding to the mean absolute error (MAE), computed over mini-batches of size 64, with the training batches shuffled at each epoch and the last incomplete batch dropped. Model training and evaluation were repeated using three different seeds to assess robustness to model initialization. This procedure was applied to all methods, and results were averaged across seeds for each participant and task. The training setup was based on previous studies and empirical tuning \cite{Mendez2021}.

\subsection{Data Analysis}
Six reference methods were implemented in this study to compare performance in simultaneous and proportional hand-movement prediction against the proposed physiologically inspired method. Specifically, the methods considered were Ridge Regression (RR), K-Nearest Neighbors (KNN), Histogram-based Gradient Boosting (HGB) Regression, Support Vector Regressor (SVR), Multilayer Perceptron (MLP), and a 2D Convolutional Neural Network (CNN), configured according to previous reports \cite{Molinari2025, Nowak2023, Mendez2021}. Detailed configurations for each method can be found in Section 3 of the \textit{Supplementary Material}. For RR, KNN, HGB, and SVR, hyperparameters were selected using a protocol similar to that followed in the study by \cite{Molinari2025}. The MLP and CNN implemented here correspond to the methods proposed and evaluated in \cite{Mendez2021}. RR, KNN, HGB, SVR, and MLP use as input nine features extracted from 100 ms windows of the sEMG signals, with 50\% overlap: RMS, mean absolute value, waveform length, standard deviation, zero crossings, maximum absolute value, logarithmic detector, mean frequency, and median frequency, with the latter two computed from the fast Fourier transform of the same window, yielding a 1152-dimensional input (9 features $\times$ 128 channels) per window. The CNN instead receives the band-pass-filtered sEMG as input over the same 100 ms window (205 samples), yielding a $1\times205\times128$ input per window. The segmentation of the data into training, validation, and test sets was kept consistent with that used in the proposed method. Likewise, the MLP and CNN followed the same training scheme as $SNN_{Physio}$, whereas RR, KNN, HGB, and SVR were fitted on the same training set using their standard procedures, ensuring a fair comparison across methods.

\subsubsection{Ablation analysis}
To quantify the contribution of each physiologically inspired component, $SNN_{Physio}$ was compared with three variants whose hyperparameters were also optimized with Optuna, following the same procedure as for $SNN_{Physio}$. The conventional SNN ($SNN_{Conv}$) \cite{GuerreroMendez2026} shared the LIF architecture and MLP readout of $SNN_{Physio}$, but used identical neurons in both layers (fixed thresholds and time constants), unconstrained synaptic weights, and the adaptive temporal encoder proposed by Sun \textit{et al.} \cite{Sun2023}, which emits a spike when the absolute difference between consecutive samples of the filtered sEMG exceeds an adaptive level (the mean plus a multiple of the standard deviation of the sample-to-sample differences). The other two variants were obtained by removing one of the two main components of $SNN_{Physio}$: $SNN_{SpkEnd}$ retained the spike encoder but used identical thresholds and time constants in the second layer, whereas $SNN_{HeteMN}$ retained the heterogeneous second layer but replaced the spike encoder with the adaptive temporal encoder. For this analysis, all 30 experimental conditions were included.

\subsubsection{Exploratory analyses}
The effects of the physiologically inspired components (Section~\ref{sec:PhysioComponents}) were explored by varying their parameters. For simplicity, these analyses were restricted to a single condition (FG task, neutral wrist position at 90$^{\circ}$, and extended elbow), whereas the ablation analysis and the comparison with baseline methods included all 30 conditions.

\begin{itemize}
    \item \textit{Spike encoder firing rate}: The effects of varying $\theta$ were explored, with values set to 25, 50, 100, 150, 200, 250, 300, 350, 400, 450, 500, and 600, which respectively generated the following firing rates of the spike encoder 8.50, 16.98, 33.82, 50.43, 66.85, 83.09, 99.16, 115.09, 130.86, 146.46, 161.87, and 192.19 Hz.

    \item \textit{Variability of input spikes}: In this study, orders of 1, 5, 15, 25, 50, 75, 100, 125, and 150 were considered, corresponding to theoretical CoV values of 100, 45, 26, 20, 14, 12, 10, 9, and 8 \%, respectively, plus a deterministic encoder (fixed threshold, CoV of 0\%).

    \item \textit{Excitation/inhibition balance}: Three E/I balance configurations were defined: 65/35, 70/30, and 80/20. Additionally, a fourth configuration without any sign constraint (WB) was evaluated, using the network's standard, unconstrained initialization.
\end{itemize}

\subsubsection{Performance and interpretability assessments}
Multiple performance metrics were employed to characterize both the effects of the physiological adaptations and the comparison with baseline methods. Specifically, the following metrics were used: root mean square error (RMSE), $R^2$, Pearson correlation coefficient ($\rho$), MAE, and maximum cross-correlation ($XC_{max}$). RMSE, $R^2$, $\rho$, and MAE were computed between the measured joint angle at the last sample of each window and the prediction obtained from that window, without any temporal shift, whereas $XC_{max}$ was computed over all lags; both procedures were applied identically to all methods. The lag at which the cross-correlation reached its maximum ($Lag_{XC}$) was also reported as an estimate of the prediction delay of each method. To obtain a global measure of performance in the simultaneous estimation of all finger joints, a variance-weighted version of these metrics, computed on the non-standardized joint angles, was adopted. This approach assigns greater importance to outputs with larger ranges of motion, which are more informative for evaluating the decoder’s ability to predict multiple degrees of freedom simultaneously. In the analyses of the physiological adaptations, only the $R^2$ and $\rho$ metrics were considered, due to their ease of interpretation and widespread use in evaluating the efficiency of decoding methods \cite{Bi2019}.

Additionally, to assess the physiological interpretability of $SNN_{Physio}$ through its internal activity, the instantaneous population firing rate, estimated as the smoothed composite spike train (sCST; the sum of the spike trains of all neurons in a layer, normalized by the number of neurons and smoothed by zero-phase filtering with a 25 ms Hann window), and several statistical properties of the inter-spike interval (ISI) distribution were computed, including the mean, CoV, skewness, and kurtosis. Population synchrony, the coherence between the sCSTs of two equally sized, non-overlapping groups of neurons from each layer, and the gain between layers were also computed (see Section 6 of the \textit{Supplementary Material}). These internal activity measures were compared across tasks in the conditions with extended elbow and 90$^{\circ}$ wrist posture, to isolate the effect of the task, since these properties may also be modulated by elbow and wrist posture.

\subsubsection{Statistical analysis}
The normality of the data distribution for each condition was assessed using the Shapiro–Wilk test. As most conditions did not meet the assumption of normality ($p < 0.05$), non-parametric statistical analyses were employed. To evaluate overall performance differences among conditions during the physiological explorations, the Friedman test was applied, considering the repeated-measures nature of the data. All tests used one value per participant (n = 23), obtained by averaging across seeds and, for the method comparison, taking the median across the 30 conditions. The hypotheses are as follows: (i) for the firing rate of the spike encoder, the alternative hypothesis states that higher firing rates lead to improved performance compared to lower firing rates; (ii) for input spike variability (input noise), the alternative hypothesis states that a reduction in noise results in a significant improvement in performance; (iii) for the E/I balance, the alternative hypothesis states that decoding performance differs across the evaluated E/I configurations; and (iv) for the comparison across methods, the alternative hypothesis states that $SNN_{Physio}$ achieves higher performance than the baseline methods. For the internal activity measures of the SNN, the alternative hypothesis states that they differ across motor tasks, with each SNN layer analyzed separately. 

The effect size for the overall analysis was assessed using Kendall’s coefficient of concordance ($W$). When significant differences were detected, post-hoc pairwise comparisons were performed using the two-sided Wilcoxon signed-rank test, with a Bonferroni correction applied to the resulting $p$-values to account for multiple comparisons; the adjusted $p$-values are reported. For the post-hoc comparisons, effect size was quantified using the rank-biserial correlation coefficient ($r_{rb}$), which measures the magnitude of the difference between paired conditions \cite{tomczak2014need}. Effect sizes were interpreted according to the following thresholds: small (0.1–$<$0.4), medium (0.4–$<$0.6), and large ($\geq$0.6) \cite{mangiafico2016summary}. All tests adopted a significance level of $p<0.05$. Descriptive values are reported as the median and interquartile range (IQR).

\section{Results}
\subsection{Data characterization}
\begin{figure}
    \centering
    \includegraphics[width=0.55\linewidth]{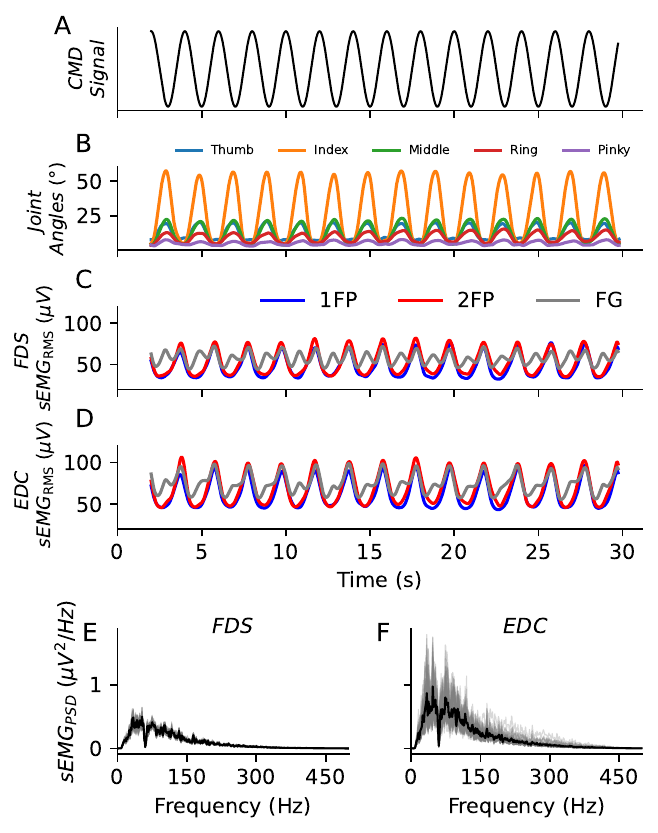}
    \caption{Example of the recordings obtained during the experiments, averaged across 23 participants, in the condition with extended elbow and 90$^{\circ}$ wrist posture. (A) Command signal used to drive the virtual hand presented through the Unity interface, which participants were instructed to follow. (B) Joint angles during the 1FP task, including the metacarpophalangeal joint of the thumb and the proximal interphalangeal joints of the index, middle, ring, and little fingers. (C) $sEMG_{RMS}$ (root mean square) from the flexor digitorum superficialis (FDS) muscle for each hand task, averaged across channels. (D) $sEMG_{RMS}$ from the extensor digitorum communis (EDC) muscle for the same tasks. (E--F) $sEMG_{PSD}$ (power spectral density) of the FDS and EDC muscles, respectively, estimated with Welch's method (1 s Hann windows without overlap; 1 Hz frequency resolution). The black line represents the average across channels, while the gray lines correspond to individual sEMG channels. 1FP: one-finger pinch; 2FP: two-finger pinch; FG: full-hand grasp.}
    \label{fig:EMGmuscles}
\end{figure}

All samples were triangulated, with a camera calibration reprojection error of 0.17 px (range: 0.13--0.25 px) across participants and a hand-landmark reprojection error of 9.1 px on average (95th percentile: 18.6 px; 720 $\times$ 540 px images). The joint angles followed the 0.5 Hz command, and the sEMG envelopes of both muscles were modulated at the same frequency (Figure~\ref{fig:EMGmuscles}).

\subsection{Ablation of the physiologically inspired components}\label{sec:Ablation}
\begin{figure}
    \centering
    \includegraphics[width=0.52\linewidth]{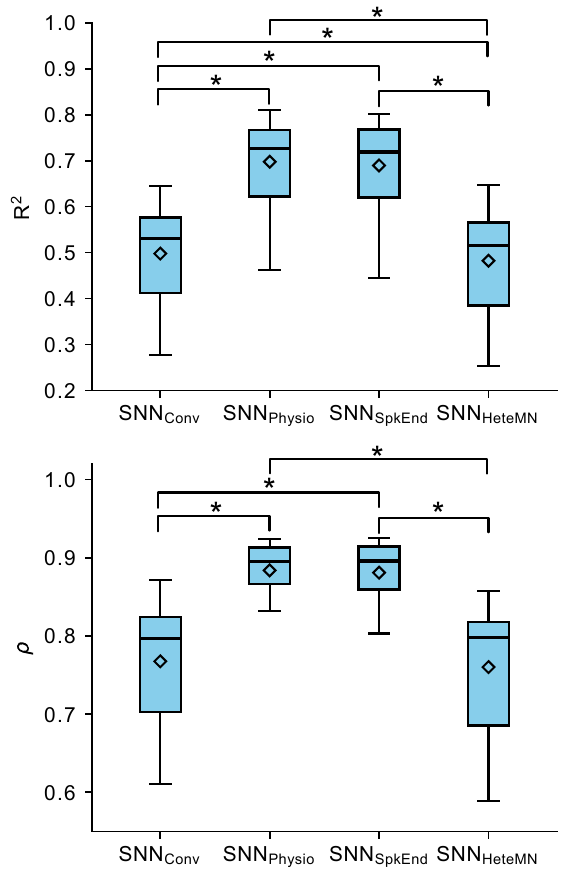}
    \caption{Ablation of the physiologically inspired components. Coefficient of determination ($R^2$; top) and Pearson correlation coefficient ($\rho$; bottom) of the proposed method ($SNN_{Physio}$) and its variants: conventional SNN ($SNN_{Conv}$), without the heterogeneous second layer ($SNN_{SpkEnd}$), and with the spike encoder replaced by the adaptive temporal encoder ($SNN_{HeteMN}$). Values correspond to the median across the 30 conditions for each participant. * denotes statistically significant differences ($p<$0.05).}
    \label{fig:Ablation}
\end{figure}

The Friedman test revealed significant differences between $SNN_{Physio}$ and its variants across all 30 experimental conditions (Figure~\ref{fig:Ablation}), with large effect sizes for both $R^2$ ($p<$ 0.0001, $W$ = 0.76) and $\rho$ ($p<$ 0.0001, $W$ = 0.73). Post-hoc analyses using the Wilcoxon signed-rank test showed that $SNN_{Physio}$ outperformed $SNN_{Conv}$ and $SNN_{HeteMN}$ for both metrics ($p<$ 0.0001, $r_{rb}\geq$ 0.99), whereas no significant differences were found between $SNN_{Physio}$ and $SNN_{SpkEnd}$ ($R^2$: $p$ = 0.80, $r_{rb}$ = 0.36; $\rho$: $p$ = 1.00, $r_{rb}$ = 0.33). Likewise, $SNN_{SpkEnd}$ outperformed both $SNN_{Conv}$ and $SNN_{HeteMN}$ ($p<$ 0.0001, $r_{rb}\geq$ 0.98). In contrast, $SNN_{HeteMN}$ showed a slightly lower $R^2$ than $SNN_{Conv}$ ($p$ = 0.02, $r_{rb}$ = 0.67), with no significant difference for $\rho$ ($p$ = 0.23, $r_{rb}$ = 0.49). In descriptive terms, the median $R^2$ was 0.73 [0.62, 0.77] for $SNN_{Physio}$, 0.72 [0.62, 0.77] for $SNN_{SpkEnd}$, 0.52 [0.39, 0.57] for $SNN_{HeteMN}$, and 0.53 [0.41, 0.58] for $SNN_{Conv}$, while the median $\rho$ was 0.89 [0.87, 0.91], 0.90 [0.86, 0.91], 0.80 [0.69, 0.82], and 0.80 [0.70, 0.82], respectively. Overall, the spike encoder accounted for the performance gain, whereas the heterogeneous second layer provided no significant additional improvement.

\subsection{Effects of the spike encoder firing rate}\label{sec:EffectsTheta}
\begin{figure}
    \centering
    \includegraphics[width=0.65\linewidth]{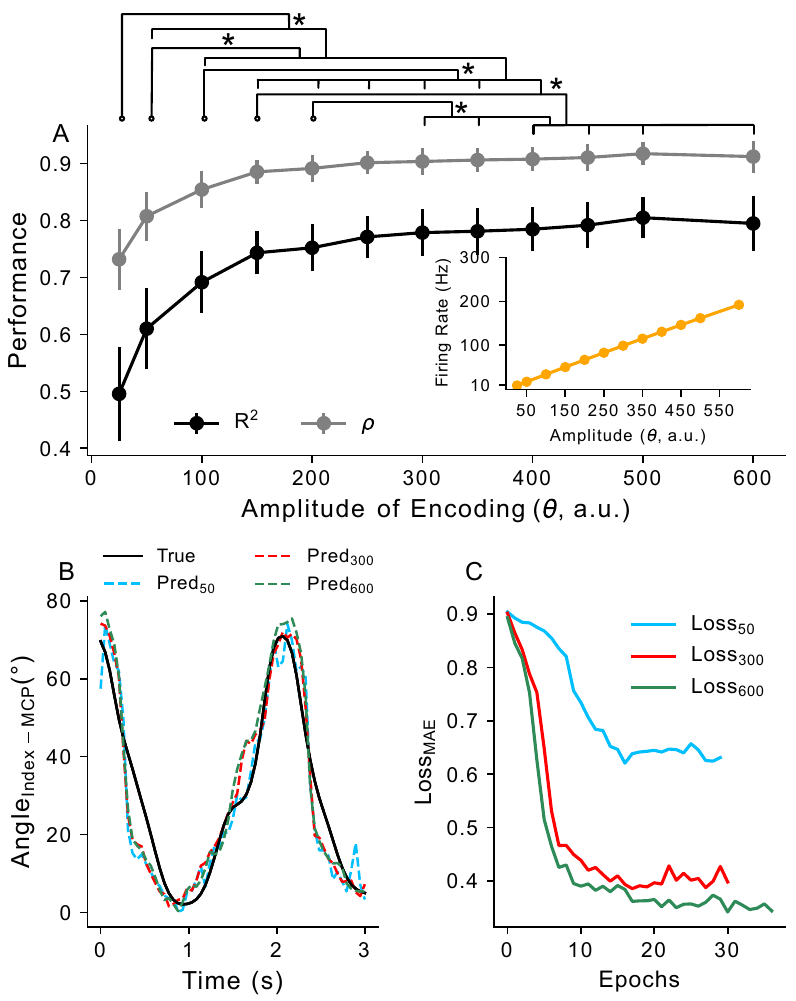}
    \caption{Effect of the amplitude of the spike encoder's modulating signal ($\theta$) on decoding performance in the full-hand grasp (FG) task (wrist at 90$^{\circ}$, elbow extended). (A) Coefficient of determination ($R^2$) and Pearson correlation coefficient ($\rho$) as a function of $\theta$, displaying the population's mean and 95\% confidence interval. The relationship between $\theta$ and the spike encoder's firing rate is depicted in the subfigure (orange line). (B) A comparison of the predictions for $\theta$ values of 50, 300, and 600 with the true signal (index finger angle at the metacarpophalangeal (MCP) joint for a single randomly selected participant). (C) The evolution of the validation L1 loss function ($Loss_{MAE}$, mean absolute error) for each of the three $\theta$ values in a single randomly selected participant during training. * denotes statistically significant differences in $R^2$ ($p<$0.05).}
    \label{fig:AmplitudePer}
\end{figure}

An increase in decoding performance was observed as the firing rate of the spike encoder increased, controlled by the amplitude $\theta$ of the modulating signal (Figure \ref{fig:AmplitudePer}; FG task, wrist at 90$^{\circ}$, elbow extended). Notice that the relationship between amplitude and firing rate is directly proportional (inset in Panel A).

The Friedman test revealed a significant and large effect for both $R^2$ ($p<$ 0.0001, $W$ = 0.69) and $\rho$ ($p<$ 0.0001, $W$ = 0.73). Post-hoc analyses using the Wilcoxon signed-rank test showed that an amplitude of 25 yielded significantly lower performance than all higher amplitudes (highest $p<$ 0.0001, $r_{rb}$ = 1.00). Similarly, an amplitude of 50 yielded significantly lower performance than higher amplitudes (highest $p<$ 0.001, $r_{rb}\geq$ 0.97). In addition, an amplitude of 100 showed significantly lower performance compared to amplitudes of 150 and above (highest $p$ = 0.03, $r_{rb}\geq$ 0.79), while an amplitude of 150 presented significantly lower performance compared to amplitudes of 400 and above (highest $p$ = 0.04, $r_{rb}\geq$ 0.77) and, for $\rho$, also compared to an amplitude of 250 ($p$ = 0.02, $r_{rb}$ = 0.80). For an amplitude of 200, significantly lower performance was observed compared to amplitudes of 300 and above for both metrics (highest $p$ = 0.03, $r_{rb}\geq$ 0.79). From an amplitude of 250 onward, performance reached a plateau. All significant differences had large effect sizes. Collectively, these results suggest that beyond an amplitude  $\theta$ of 250, no consistent further improvements were observed in either metric.

In line with the above, and given that performance reached a plateau for amplitudes of 250 and above, the $R^2$ for this condition reached a value of 0.77 [0.75, 0.81], while $\rho$ was 0.90 [0.89, 0.93]. Additionally, Panels B and C illustrate the less accurate predictions of the SNN when a low amplitude is used ($\theta$ = 50), which also reached a higher validation loss during training. In contrast, amplitudes of 300 and 600 yielded more accurate predictions and reached similar validation losses.

\subsection{Effects of the variability of input spikes}
\begin{figure}
    \centering
    \includegraphics[width=0.65\linewidth]{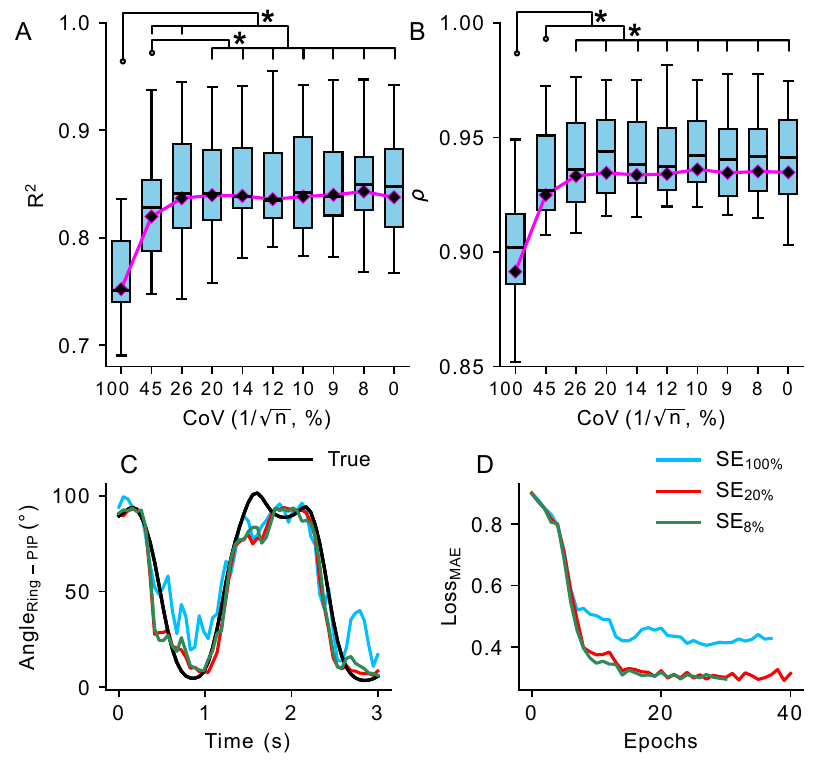}
    \caption{Effects of spike encoder order (input noise measured as the coefficient of variation, CoV) on decoding performance in the full-hand grasp (FG) task (wrist at 90$^{\circ}$, elbow extended). (A–B) Distribution of the coefficient of determination ($R^2$) and Pearson correlation coefficient ($\rho$) for the population group, respectively. (C) A temporal comparison of the predictions and the true signal for a single randomly selected participant's ring finger proximal interphalangeal (PIP) joint angle under three spike encoder (SE) orders ($SE_{x}$ digits represent the CoV). (D) A single randomly selected participant's evolution of the L1 loss function ($Loss_{MAE}$, mean absolute error) during training for the $SE_{x}$. * denotes statistically significant differences ($p<$0.05).}
    \label{fig:OrdenPer}
\end{figure}

Decoding performance was lower with the Poisson-like encoder (order 1, CoV of 100\%) than with the more regular inputs, which are obtained by increasing the order of the spike encoder (Figure \ref{fig:OrdenPer}; FG task, wrist at 90$^{\circ}$, elbow extended). The Friedman test confirmed significant differences for both $R^2$ ($p<$ 0.0001, $W$ = 0.38) and $\rho$ ($p<$ 0.0001, $W$ = 0.40), with small-to-medium effect sizes.

The Wilcoxon signed-rank post-hoc test confirmed that the spike encoder with an order of 1 (CoV of 100\%) resulted in significantly lower performance compared with all other spike encoder CoV levels, for both $R^2$ (highest $p<$ 0.0001, $r_{rb}\geq$ 0.99) and $\rho$ ($p<$ 0.0001, $r_{rb}\geq$ 0.99). Likewise, an order of 5 (CoV of 45\%) showed significantly lower performance than all more regular inputs (CoV $\leq$ 26\%) for $\rho$ (highest $p$ = 0.02, $r_{rb}\geq$ 0.78), and than CoV levels of 20\% and below for $R^2$ (highest $p$ = 0.03, $r_{rb}\geq$ 0.76). In contrast, no significant differences were found among CoV levels of 26\% and below, including the deterministic encoder (CoV of 0\%) ($p$ = 1.00, $r_{rb}\leq$ 0.51). In descriptive terms, the median $R^2$ was 0.75 [0.74, 0.80] for a CoV of 100\%, 0.83 [0.79, 0.85] for 45\%, and ranged from 0.84 to 0.85 for CoV levels of 26\% and below, while the median $\rho$ was 0.90 [0.89, 0.92], 0.93 [0.92, 0.95], and approximately 0.94, respectively.

The above results were also reflected in the SNN’s ability to predict the target signal and in its learning process (Panels C and D), where a CoV of 100\% resulted in higher validation loss during training.

\subsection{Excitation/inhibition balance}
\begin{figure}
    \centering
    \includegraphics[width=0.65\linewidth]{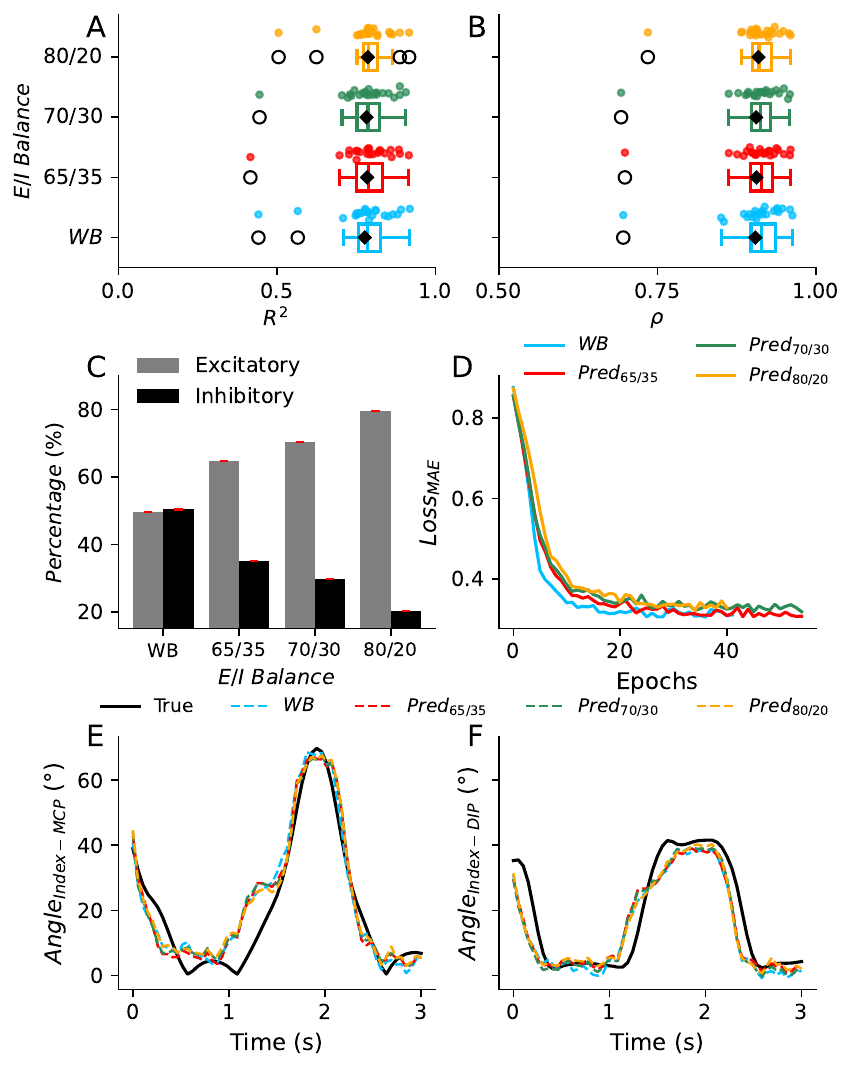}
    \caption{Effect of excitation/inhibition (E/I) balance on decoding performance in the full-hand grasp (FG) task (wrist at 90$^{\circ}$, elbow extended). WB indicates there was no restriction in the balance for the initialization and training of synaptic weights. (A) Coefficient of determination ($R^2$) for the different E/I configurations. (B) Pearson correlation coefficient ($\rho$) for the different E/I configurations. (C) Mean and 95\% confidence intervals of the percentage of excitatory (positive) and inhibitory (negative) first-layer synaptic weights after training. (D) The evolution of the L1 loss function ($Loss_{MAE}$, mean absolute error) during training for a single randomly selected participant under each balance condition. (E–F) Metacarpophalangeal (MCP) and distal interphalangeal (DIP) joint angles of the index finger for each E/I balance, showing predictions and the ground-truth signal for a single randomly selected participant.}
    \label{fig:BalancesPer}
\end{figure}

Decoding performance was not significantly affected by the E/I balance imposed during training (Figure \ref{fig:BalancesPer}; FG task, wrist at 90$^{\circ}$, elbow extended), as the Friedman test revealed no significant differences between conditions for either $R^2$ ($p$ = 0.50, $W$ = 0.03) or $\rho$ ($p$ = 0.40, $W$ = 0.04). After training, the proportions of excitatory and inhibitory first-layer weights matched the imposed configurations across participants (Panel C), as expected from the sign constraint, whereas in the unconstrained configuration (WB) they remained close to 50\% each. Panels D–F illustrate the loss function and the predicted joint angles for a representative participant for each E/I balance condition, supporting the small effect of this property on kinematic prediction.

\begin{figure}[ht]
    \centering
    \includegraphics[width=0.95\linewidth]{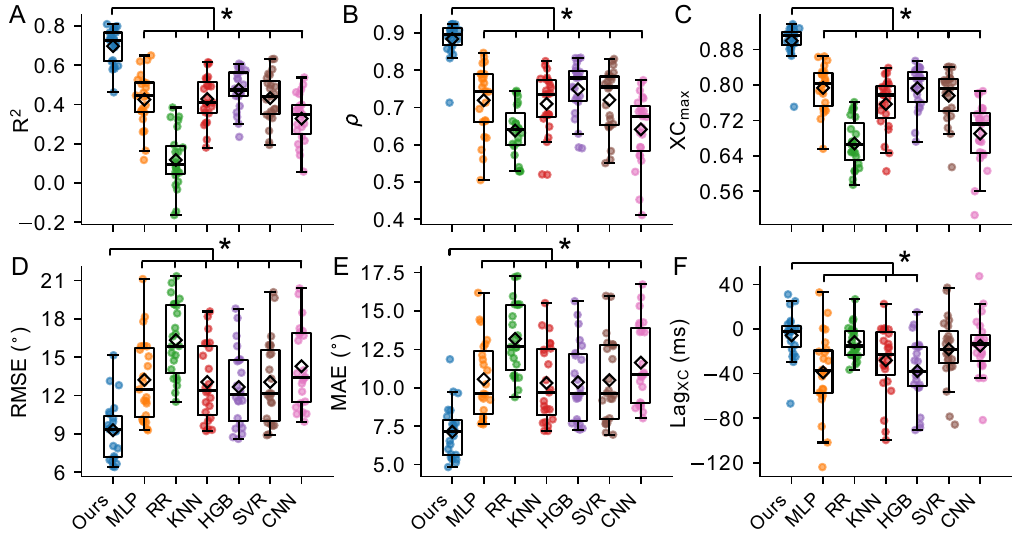}
    \caption{Comparison of the proposed method ($SNN_{Physio}$, labeled Ours) performance with all baseline methods. Each panel represents one of the evaluated performance metrics: (A) coefficient of determination ($R^2$), (B) Pearson correlation coefficient ($\rho$), (C) maximum cross-correlation ($XC_{max}$), (D) root mean square error (RMSE), (E) mean absolute error (MAE), and (F) cross-correlation lag ($Lag_{XC}$). Values correspond to the median across the 30 conditions for each participant. * indicates statistically significant differences between $SNN_{Physio}$ and the baseline methods indicated by each bracket. MLP: multilayer perceptron; RR: ridge regression; KNN: k-nearest neighbors; HGB: histogram-based gradient boosting; SVR: support vector regressor; CNN: convolutional neural network.}
    \label{fig:Comparison}
\end{figure}

\subsection{Performance comparison}
The ability of $SNN_{Physio}$ to simultaneously and proportionally predict finger kinematics demonstrated significantly superior performance compared to the implemented reference methods across all 30 experimental conditions (Figure \ref{fig:Comparison} and Table~\ref{tab:Comparison}). The Friedman test revealed a significant and large effect on performance ($p < 0.0001$, $W \geq$ 0.79) for $R^2$, $\rho$, $XC_{max}$, RMSE, and MAE. These findings were confirmed through a post-hoc analysis using the Wilcoxon signed-rank test. Specifically, for the $R^2$, $\rho$, and $XC_{max}$ metrics (Figure~\ref{fig:Comparison}A--C), large and significant differences were observed when comparing $SNN_{Physio}$ with MLP ($p < 0.0001$, $r_{rb} \geq$ 0.99), RR ($p < 0.0001$, $r_{rb}$ = 1.00), KNN ($p < 0.0001$, $r_{rb}$ = 1.00), HGB ($p < 0.0001$, $r_{rb} \geq$ 0.99), SVR ($p < 0.0001$, $r_{rb}$ = 1.00), and CNN ($p < 0.0001$, $r_{rb}$ = 1.00), with consistent patterns in both $p$-values and effect sizes across metrics. For RMSE and MAE metrics (Figure~\ref{fig:Comparison}D--E), the Wilcoxon signed-rank post-hoc test revealed large and significant differences when comparing $SNN_{Physio}$ with MLP ($p < 0.0001$, $r_{rb}$ = 1.00), RR ($p < 0.0001$, $r_{rb}$ = 1.00), KNN ($p < 0.0001$, $r_{rb}$ = 1.00), HGB ($p < 0.0001$, $r_{rb}$ = 1.00), SVR ($p < 0.0001$, $r_{rb}$ = 1.00), and CNN ($p < 0.0001$, $r_{rb}$ = 1.00).

\begin{table}[ht]
    \centering
    \caption{Performance of $SNN_{Physio}$ and the baseline methods. Values are the median (top) and interquartile range (IQR; bottom) across participants (n = 23), computed from the median across the 30 conditions for each participant. $R^2$: coefficient of determination; $\rho$: Pearson correlation coefficient; $XC_{max}$: maximum cross-correlation; RMSE: root mean square error; MAE: mean absolute error; $Lag_{XC}$: cross-correlation lag; MLP: multilayer perceptron; RR: ridge regression; KNN: k-nearest neighbors; HGB: histogram-based gradient boosting; SVR: support vector regressor; CNN: convolutional neural network.}
    \label{tab:Comparison}
    \begin{tabular}{@{}lcccccc@{}}
        \toprule
        Method & $R^2$ & $\rho$ & $XC_{max}$ & RMSE ($^{\circ}$) & MAE ($^{\circ}$) & $Lag_{XC}$ (ms) \\
        \midrule
        $SNN_{Physio}$ (Ours) & 0.73 & 0.89 & 0.91 & 9.36 & 7.15 & $-$2 \\
         & [0.62, 0.77] & [0.87, 0.91] & [0.89, 0.92] & [7.18, 10.38] & [5.65, 7.92] & [$-$16, 2] \\
        \addlinespace
        MLP & 0.45 & 0.74 & 0.80 & 12.48 & 9.61 & $-$37 \\
         & [0.36, 0.51] & [0.66, 0.79] & [0.75, 0.83] & [10.29, 15.72] & [8.32, 12.40] & [$-$58, $-$19] \\
        \addlinespace
        RR & 0.09 & 0.64 & 0.67 & 15.87 & 12.66 & $-$15 \\
         & [0.05, 0.19] & [0.60, 0.69] & [0.63, 0.71] & [13.74, 19.10] & [11.17, 15.41] & [$-$23, $-$2] \\
        \addlinespace
        KNN & 0.41 & 0.74 & 0.78 & 12.37 & 9.70 & $-$23 \\
         & [0.36, 0.51] & [0.67, 0.77] & [0.73, 0.80] & [10.47, 15.91] & [8.24, 12.51] & [$-$42, $-$3] \\
        \addlinespace
        HGB & 0.47 & 0.78 & 0.81 & 12.09 & 9.62 & $-$38 \\
         & [0.44, 0.56] & [0.72, 0.80] & [0.76, 0.83] & [10.00, 14.78] & [7.86, 12.18] & [$-$51, $-$16] \\
        \addlinespace
        SVR & 0.44 & 0.76 & 0.79 & 12.16 & 9.64 & $-$19 \\
         & [0.35, 0.52] & [0.65, 0.78] & [0.74, 0.81] & [10.04, 15.59] & [7.95, 12.76] & [$-$31, $-$2] \\
        \addlinespace
        CNN & 0.35 & 0.67 & 0.71 & 13.44 & 10.86 & $-$13 \\
         & [0.25, 0.40] & [0.58, 0.70] & [0.65, 0.74] & [11.47, 16.88] & [9.01, 13.88] & [$-$29, $-$6] \\
        \bottomrule
    \end{tabular}
\end{table}

The Friedman test also revealed a significant effect ($p < 0.0001$) for the cross-correlation lag ($Lag_{XC}$; Figure~\ref{fig:Comparison}F), with a medium effect size ($W$ = 0.52). The median lag of $SNN_{Physio}$ was close to zero ($-$2 ms), whereas the baseline methods showed negative lags (medians between $-$38 and $-$13 ms; Table~\ref{tab:Comparison}), indicating that their predictions tended to precede the measured joint angles. The lag of $SNN_{Physio}$ differed significantly from those of MLP ($p$ = 0.009, $r_{rb}$ = 0.79), KNN ($p$ = 0.005, $r_{rb}$ = 0.82), and HGB ($p < 0.0001$, $r_{rb}$ = 0.97), whereas no significant differences were found with RR, SVR, and CNN. However, the significant differences were smaller than the 50 ms step between consecutive predictions.

Regarding model complexity, $SNN_{Physio}$ required 29,903 trainable parameters, fewer than the MLP (337,295) and the CNN (1,405,391), and more than the RR (17,295), whereas the sizes of the KNN, HGB, and SVR methods depend on the training data.

Statistical results were reported only for comparisons involving $SNN_{Physio}$, in accordance with the study's objective and scope. In Section 5 of the \textit{Supplementary Material}, additional comparisons were provided between $SNN_{Physio}$ and the reference method that achieved the highest $R^2$ (HGB) across all 30 experimental conditions and for each individual joint. These complementary results showed statistically significant improvements in most task–joint combinations, particularly during the FG task, further supporting the superior performance of $SNN_{Physio}$.

\subsection{Interpretability of the SNN internal activity}
\begin{figure}
    \centering
    \includegraphics[width=0.65\linewidth]{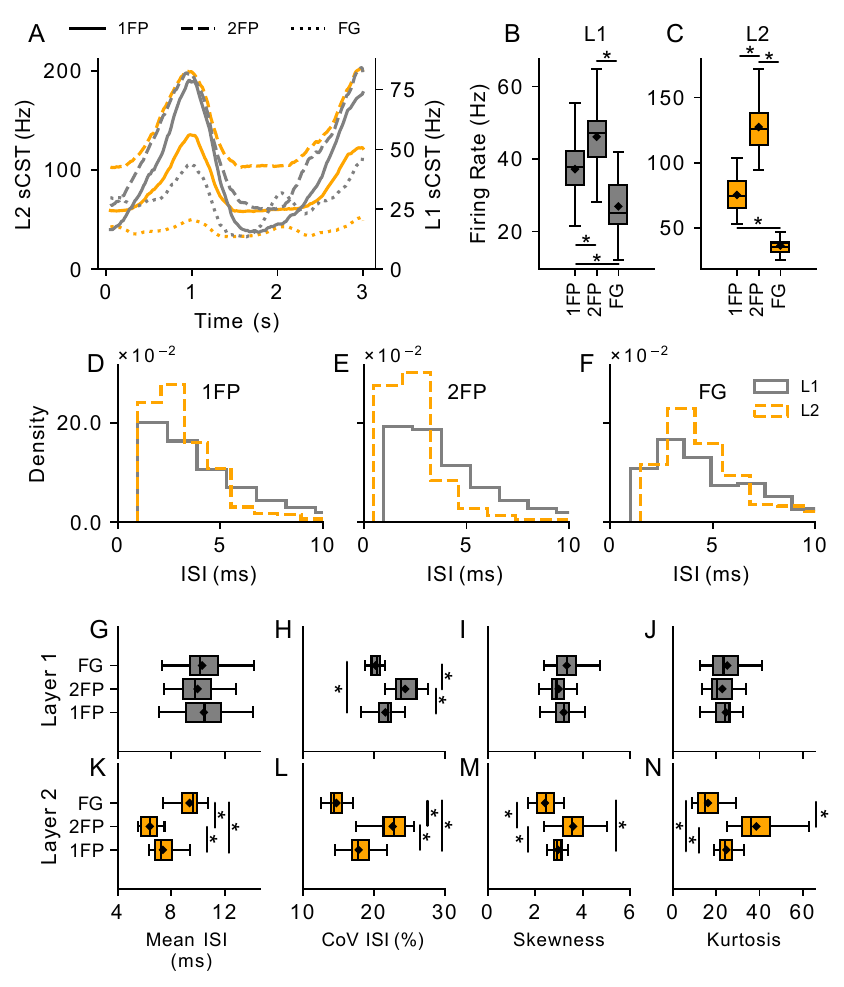}
    \caption{Firing properties of $SNN_{Physio}$ as a function of task (one-finger pinch, 1FP; two-finger pinch, 2FP; and full-hand grasp, FG) in the conditions with extended elbow and 90$^{\circ}$ wrist posture. (A) Smoothed composite spike train (sCST) of layer 1 (L1, gray, right axis) and layer 2 (L2, orange, left axis) over time in the test set for each task (line style), averaged across participants. (B–C) Boxplots of population firing rate by task for layer 1 and layer 2, respectively. (D–F) Inter-spike interval (ISI) distributions for each task for a representative participant, separated by layer (L1: solid gray; L2: dashed orange). (G–J) Statistical properties of the ISIs for layer 1: mean ISI, coefficient of variation (CoV), skewness, and kurtosis, respectively. (K–N) Corresponding statistical properties for layer 2. In (B–C) and (G–N), boxplots summarize one value per participant (n = 23), averaged across neurons. * denotes statistically significant differences ($p<$0.05).}
    \label{fig:Neuronal}
\end{figure}

The firing properties of $SNN_{Physio}$ in the conditions with extended elbow and 90$^{\circ}$ wrist posture differed across motor tasks in both layers (Figure \ref{fig:Neuronal}). The Friedman test revealed significant differences in firing rate (Figure~\ref{fig:Neuronal}A--C) across tasks in layer 1 ($p<$ 0.0001, $W$ = 0.58) and layer 2 ($p<$ 0.0001, $W$ = 0.96). Post-hoc analyses using the Wilcoxon signed-rank test showed that 2FP produced the highest firing rates in both layers, followed by 1FP and FG, with significant differences between all pairs of tasks in layer 1 (highest $p$ = 0.002, $r_{rb}\geq$ 0.78) and layer 2 ($p<$ 0.0001, $r_{rb}\geq$ 0.99).

In all tasks and layers, the ISI distributions were positively skewed and showed high kurtosis, with a pronounced concentration of short intervals and a long tail toward longer ISIs (Figure~\ref{fig:Neuronal}D--F). In layer 1 (Figure~\ref{fig:Neuronal}G--J), only the CoV differed across tasks (Friedman $p<$ 0.0001, $W$ = 0.76), being highest for 2FP, followed by 1FP and FG (highest $p$ = 0.01, $r_{rb}\geq$ 0.67), whereas the mean ISI, skewness, and kurtosis did not differ significantly (Friedman $p\geq$ 0.12, $W\leq$ 0.09). In layer 2 (Figure~\ref{fig:Neuronal}K--N), all ISI statistics differed across tasks (Friedman $p<$ 0.0001, $W$ between 0.69 and 0.96), with significant differences between all pairs of tasks (highest $p<$ 0.001, $r_{rb}\geq$ 0.88): 2FP showed the shortest mean ISIs and the highest CoV, skewness, and kurtosis, followed by 1FP and FG.

For conciseness, the additional population measures are presented in Section 6 of the \textit{Supplementary Material}: synchrony, the coherence between the sCSTs of two equally sized, non-overlapping groups of neurons from each layer, and the gain between layers. These measures also differed across tasks, except for the sCST coherence of layer 1.

\section{Discussion}
In the present study, we proposed a spinal-inspired SNN that significantly improved the simultaneous and proportional prediction of finger kinematics from HD sEMG signals compared with traditional machine-learning and deep-learning approaches. Notably, this improvement was achieved with substantially fewer trainable parameters than the MLP and CNN, suggesting that it is not attributable to a larger model size, in line with recent evidence that shallow SNNs can decode finger-level motor output with a small memory footprint \cite{Baracat2026}. The results showed that incorporating neurophysiological principles enhanced the biological plausibility of the architecture without compromising decoding performance. In particular, the more physiologically plausible spike-encoding model significantly improved decoding performance and was the main determinant of the improved performance of the proposed method, whereas the phenomenological representation of Henneman's size principle and the different E/I balances imposed on the synaptic weights did not significantly alter this performance. Joint kinematics were evaluated across fifteen finger angles (three joint angles per finger), enabling the simultaneous and proportional estimation of the kinematics of all fingers. The proposed method showed significant improvements across both elbow postures, all wrist positions, and all three motor tasks (see Figure 3 of the \textit{Supplementary Material}).

The problem of decoding kinematics with multiple degrees of freedom has recently gained considerable attention due to its potential applications in motor rehabilitation, making it an active area of research. Various methods have been proposed, primarily data-driven approaches such as deep learning \cite{Mendez2021, Ameri2019, Hajian2022}. However, most of these methods have been evaluated on a limited number of degrees of freedom. Other approaches have been developed for the simultaneous and proportional decoding of multiple degrees of freedom, showing improved performance over more conventional methods \cite{Simpetru2023, Smpetru2024}. Although these methods have demonstrated effectiveness in predicting movements in patients with reduced muscular activity \cite{Simpetru2023}, they still present several computational and methodological challenges. For example, their incorporation into personalized rehabilitation sessions may be compromised due to the lack of neurophysiological outputs that these methods can generate; the large amount of data required for proper performance, where data augmentation strategies have been implemented \cite{Yang2022}, which may introduce biases in the data or signal without an adequate physiological representations \cite{Halmich2025}; or the limited performance that these methods may present under specific conditions, such as posture or limb position \cite{GuerreroMndez2023}.

Although the use of classical architectures is not inherently incorrect, we hypothesize that methods with more physiological representations and a closer correspondence to the structures of the central nervous system may improve performance in decoding multiple degrees of freedom across different conditions (postures, positions, and tasks). Additionally, such methods may produce more physiologically meaningful outputs that make the decoding interpretable, much as motor unit firing properties provide insight into the neural control of impaired muscles \cite{Negro2020, Duan2025}. In this context, as an alternative, the use of an SNN --- whose architecture inherently relies on computational models such as LIF neurons, which have been widely used to study neuronal dynamics \cite{Reyes2026} --- allows the incorporation of physiological realism by including dynamics of membrane potential integration and synaptic current similar to those observed in biological neurons, such as spinal motor neurons \cite{Elias2012}. Therefore, incorporating these models, together with other physiologically grounded representations, may have a significant impact on the next generation of physiology-informed neural networks.

The ablation analysis indicated that the spike encoder was the main contributor to decoding performance, as replacing it reduced the median $R^2$ by approximately 0.20, regardless of whether the second layer had heterogeneous or identical neurons (Section~\ref{sec:Ablation}). This is likely because the two encoders pass different information to the network. The proposed spike encoder converts the sEMG envelope, which is related to the neural drive to the muscle \cite{Farina2014}, into a firing rate, much as a motor neuron pool conveys this drive through its cumulative discharge rate \cite{Farina2015}. The adaptive temporal encoder, in contrast, is a threshold-based temporal-contrast method that encodes changes between consecutive samples rather than the signal level \cite{Petro2020}, so the network can only infer the activation level from how often spikes occur within each integration window. Conversely, the heterogeneous second layer did not significantly change decoding performance when combined with the spike encoder, and slightly reduced $R^2$ when combined with the adaptive temporal encoder. In motor neuron pools, recruitment thresholds that increase across neurons allow the pool to encode a wide range of input intensities, since higher-threshold neurons are recruited only as the input increases \cite{Henneman1965, Fuglevand1993}; similarly, heterogeneous time constants improve learning in SNNs, particularly in tasks with rich temporal structure \cite{PerezNieves2021}. Although the heterogeneous thresholds and time constants of the second layer phenomenologically follow these principles, they did not translate into a measurable decoding benefit. This may be partly because, in the LIF model used here, assigning a heterogeneous threshold to each neuron is equivalent to multiplying its input weights and bias by a neuron-specific factor, so a network with identical thresholds can, in principle, learn equivalent heterogeneous effective thresholds through training. Thus, the contribution of this component lies mainly in the biological plausibility of the SNN: it provides the second layer with a graded organization analogous to that of spinal motor neuron pools without compromising decoding performance, which may facilitate the interpretation of the network's internal activity in relation to spinal cord physiology.

To analyze the effects of the neurophysiological elements incorporated into the network, we evaluated the prediction performance of each element, using the variance-weighted metrics as a reference. In particular, decoding performance reached a plateau when the spike encoder inputs to the SNN exhibited a firing rate greater than or equal to approximately 83 Hz ($\theta$ = 250) (Figure \ref{fig:AmplitudePer}). It should be noted that these encoder rates exceed the discharge rates typically observed in human motor units during voluntary contractions \cite{Moritz2005, Tanzarella2020}; thus, $\theta$ should be interpreted as a gain that sets the input drive to the network rather than as a physiologically calibrated firing rate. The absence of consistent further improvements above this level suggests that, once sufficient input drive is provided, performance is no longer limited by the input. 

Subsequently, we demonstrated that fully noisy inputs to the proposed SNN, such as those imposed by a CoV of 100\% (order 1 in the spike encoder), yielded significantly lower performance than more regular inputs. Moreover, the deterministic encoder (CoV of 0\%) performed similarly to moderately regular encoders (CoV $\leq$ 26\%). Thus, input noise did not improve decoding; rather, highly irregular spike trains likely degraded the rate information available within each integration window. This contrasts with studies in which some degree of stochasticity benefits neural computation, including stochastic facilitation in neural systems \cite{McDonnell2011} and improved generalization and robustness of SNNs with noisy neuron dynamics \cite{Ma2023}. In the present method, however, noise was introduced only as variability of the input spike trains, rather than being exploited by the neuron dynamics or the learning rule \cite{Ma2023}. Importantly, decoding performance was preserved for theoretical CoV values between 14 and 26\%, which lie within the discharge variability reported for human hand motor units during steady voluntary contractions (approximately 13--30\%) \cite{Moritz2005}. This indicates that physiologically realistic levels of spike-train irregularity can be incorporated at the network input without compromising decoding performance.

Furthermore, imposing different E/I weight ratios, including more excitatory regimes (up to 80/20), did not significantly affect decoding performance (Figure \ref{fig:BalancesPer}). This result indicates that biologically plausible E/I constraints, such as the predominantly excitatory regimes widely present in the central nervous system \cite{Sahara2012}, can be incorporated without compromising prediction performance. However, the exploration of balance effects in physiological networks, particularly for understanding the input–output properties of neurons, remains an open research area \cite{Reyes2026}. 

In line with these findings, the proposed SNN enabled the characterization of neuronal activity properties, which can be regarded as interpretable latent descriptors of the proposed SNN, since their absolute values did not match physiological ranges (see below). In this sense, the proposed method addresses interpretability rather than explainability: whereas explainability methods provide post-hoc explanations of black-box models, interpretable models have internal variables that are meaningful by construction \cite{Rudin2019}, as is the case for the firing rates and ISI statistics of the SNN. In particular, the firing rate and ISI statistics (Figure~\ref{fig:Neuronal}), as well as the synchrony and sCST coherence (Section 6 of the \textit{Supplementary Material}), differed significantly across motor tasks, particularly in layer 2, indicating that the network firing properties depended on the decoded task. Moreover, the sCST coherence increased with the number of pooled neurons and was concentrated at low frequencies (Section 6 of the \textit{Supplementary Material}). This pattern is consistent with the common synaptic input that characterizes motor neuron pools \cite{Negro2011, Farina2015}, which, in the proposed SNN, arises because all neurons receive the same encoded sEMG channels. Nevertheless, since the hyperparameters were optimized independently for each task and the tasks differ intrinsically in their muscle activation patterns, these differences reflect both the task and the configuration that best decodes it, two effects that cannot be fully disentangled in the present analysis. Likewise, this analysis was restricted to a single elbow and wrist posture; how posture modulates the internal activity of the SNN was beyond the scope of this study and will be investigated in future work.

These internal descriptors could also be relevant for rehabilitation. Duan et al. \cite{Duan2025} estimated motor neuron intrinsic properties in vivo in individuals with incomplete spinal cord injury and found a longer refractory period in the tibialis anterior muscle than in non-disabled controls. The proposed method could complement such approaches by providing population-level descriptors directly from the sEMG envelope during dynamic finger movements, provided that it is further informed by the physiological mechanisms that shape motor neuron firing, such as refractoriness and spinal inhibitory circuits. Whether these descriptors reflect motor impairment in clinical populations remains to be tested.

However, despite incorporating physiologically inspired principles, the proposed SNN's firing rates were substantially higher than those typically reported experimentally for human motor neurons \cite{Moritz2005, Tanzarella2020}, particularly in layer 2 during the 2FP task. Although the SNN effectively captured the qualitative aspects of neuronal firing properties, its absolute magnitudes limit direct biological interpretation. This suggests that the improvement in decoding performance achieved with the physiologically inspired architecture did not necessarily correlate with a proportional increase in quantitative biological realism. To enhance the SNN's validity, incorporating spinal mechanisms—such as an explicit refractory period (e.g., informed by in vivo estimates \cite{Duan2025}) and recurrent or reciprocal inhibition \cite{Dernoncourt2025}—could help align neuronal dynamics with experimentally observed ranges, resulting in a more biologically plausible representation of spinal motor control.

Despite the encouraging results, some limitations should be acknowledged. For instance, the exploration was conducted using a controlled set of tasks and postures; therefore, the method's robustness under more dynamic conditions and in populations with severe motor impairments still needs to be validated. Also, the study employed a small number of electrode grids (two) for HD sEMG recordings. The grids were placed over the flexor and extensor muscles of the forearm, specifically the FDS and EDC. However, other studies have evaluated computational methods by adding two additional electrode matrices near the wrist \cite{Smpetru2024}, which may be beneficial for use in individuals with stroke or spinal cord injury. Nevertheless, for transradial amputees, our experimental approach, based on the remaining flexor and extensor musculature, may offer an advantage.

\section*{Conflict of interest}
The authors declare that they have no known competing financial interests or personal relationships that could have appeared to influence the work reported in this paper.

\section*{Acknowledgments}
We would like to thank Jose Osmar Alves-Filho and Maria Jose Granados Rodriguez for their help with experiments, and Flavio Renato Santos and Nelio Hilton Gava of the Main Design Lab at the Center for Biomedical Engineering, University of Campinas (Brazil), for their technical support in designing the experimental apparatus. We also thank SciDraw.io (https://scidraw.io/) for providing neuromuscular and human drawings used in Figure~\ref{fig:Model}A.

\section*{Funding}
This work is part of the MYOREHAB consortium (4th EU-LAC Joint Call in STI 2022), supported with funds from the São Paulo Research Foundation (FAPESP, proc. no. 22/05104-8). This work was also supported by the Brazilian Institute of Data Science (BI0S), a FAPESP Applied Research Center in Artificial Intelligence (proc. no. 20/09838-0). Cristian David Guerrero-Mendez holds a Doctoral Scholarship from FAPESP under process no. 25/03022-2; and Leonardo Abdala Elias is a Research Productivity Fellow of CNPq (Brazilian NSF, proc. no. 316320/2023-4).

\section*{Author contributions}
Cristian David Guerrero-Mendez: Writing -- original draft, Conceptualization, Formal analysis, Data curation, Methodology. Leonardo Abdala Elias: Conceptualization, Writing -- review \& editing, Supervision, Methodology, Validation, Funding acquisition, Project administration.

\section*{Data availability}
The metadata and a description of the dataset are available in the Research Data Repository of UNICAMP (REDU) \cite{Guerrero2026}; the data will be made available upon formal request. The code of the proposed method will be made available at \url{https://github.com/Cristian18042/Spinal-InspiredSNN} upon acceptance of this work.

\section*{Supplementary material}
The supplementary material is appended after the references and includes the LIF neuron model, spike encoding properties, configurations of the baseline methods, hyperparameter tuning, full performance comparison across conditions, and complementary neural analysis.

\bibliographystyle{unsrtnat_names}
\bibliography{../References}

\clearpage
\setcounter{section}{0}
\setcounter{figure}{0}
\setcounter{table}{0}
\setcounter{algorithm}{0}
\renewcommand{\thesection}{S\arabic{section}}
\renewcommand{\thefigure}{S\arabic{figure}}
\renewcommand{\thetable}{S\arabic{table}}
\renewcommand{\thealgorithm}{S\arabic{algorithm}}
\renewcommand{\theHsection}{S\arabic{section}}
\renewcommand{\theHfigure}{S\arabic{figure}}
\renewcommand{\theHtable}{S\arabic{table}}
\renewcommand{\theHalgorithm}{S\arabic{algorithm}}
\pdfbookmark[0]{Supplementary Material}{supplementary}
\begin{center}
{\LARGE\bfseries Supplementary Material}
\end{center}

\section{LIF Neuron}
\begin{figure}[ht]
    \centering
    \includegraphics[width=0.7\linewidth]{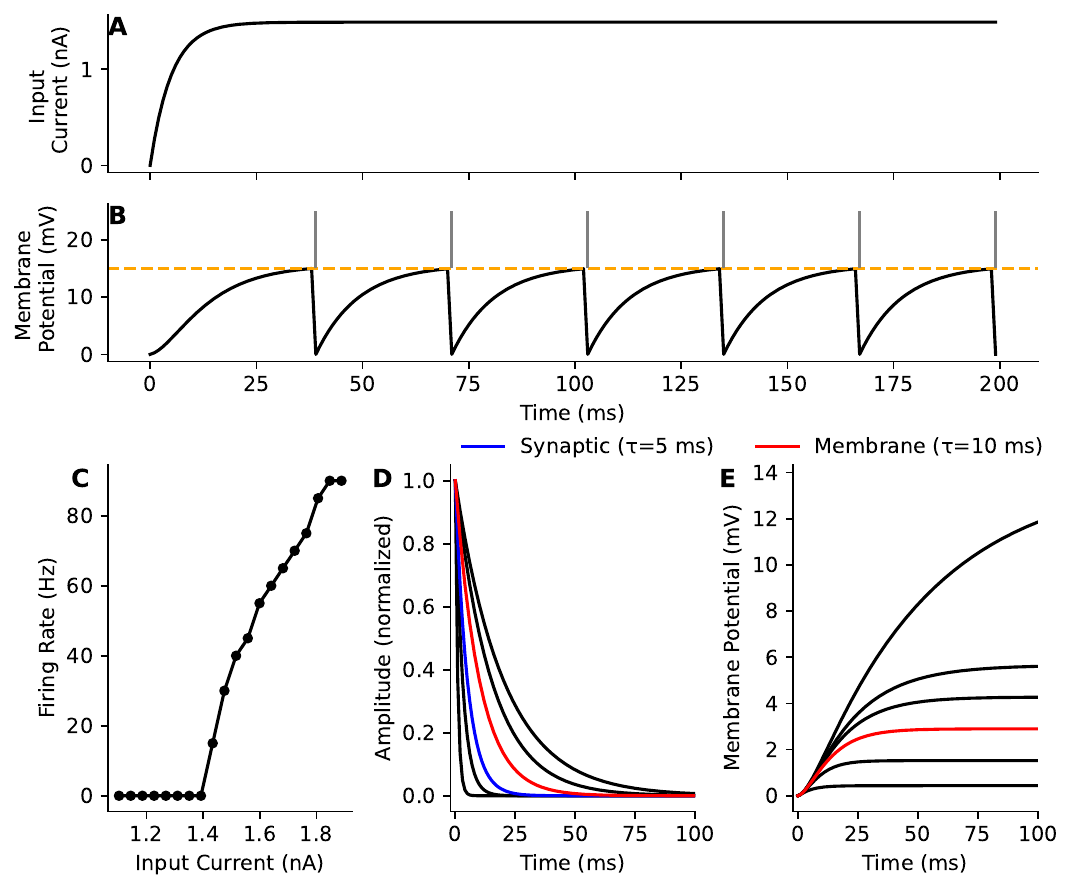}
    \caption{Dynamics of the LIF neuron model. (A) Injected input current of approximately 1.45 nA. (B) Spike train generated over a 200 ms simulation under constant stimulation. (C) Relationship between firing frequency and input current (f–I curve). (D) Normalized temporal profiles of the synaptic current and membrane potential. (E) Time course of the membrane potential.}
    \label{fig:LIFneuron}
\end{figure}

The characteristics of the LIF neuron model are shown in Figure \ref{fig:LIFneuron}. The model does not explicitly generate spikes; instead, it records the time at which the membrane potential reaches a threshold (15 mV), after which it is reset to 0 mV. In this illustrative simulation, the neuron was driven by a constant input current of approximately 1.45 nA.

The frequency–current (f–I) curve (panel C) shows an approximately proportional increase in firing rate with input current, reflecting the absence of an explicit refractory period. The normalized temporal responses indicate that synaptic current evolves faster than the membrane potential, due to its smaller time constant. In contrast, the membrane potential exhibits slower and smoother dynamics, consistent with its role as a temporal integrator. The membrane time constant modulates this integration process (panel E): smaller values lead to faster but lower-amplitude charging, whereas larger values integrate the input over longer periods, producing slower but larger depolarization.

\section{Spike encoding properties}
\begin{figure}
    \centering
    \includegraphics[width=0.8\linewidth]{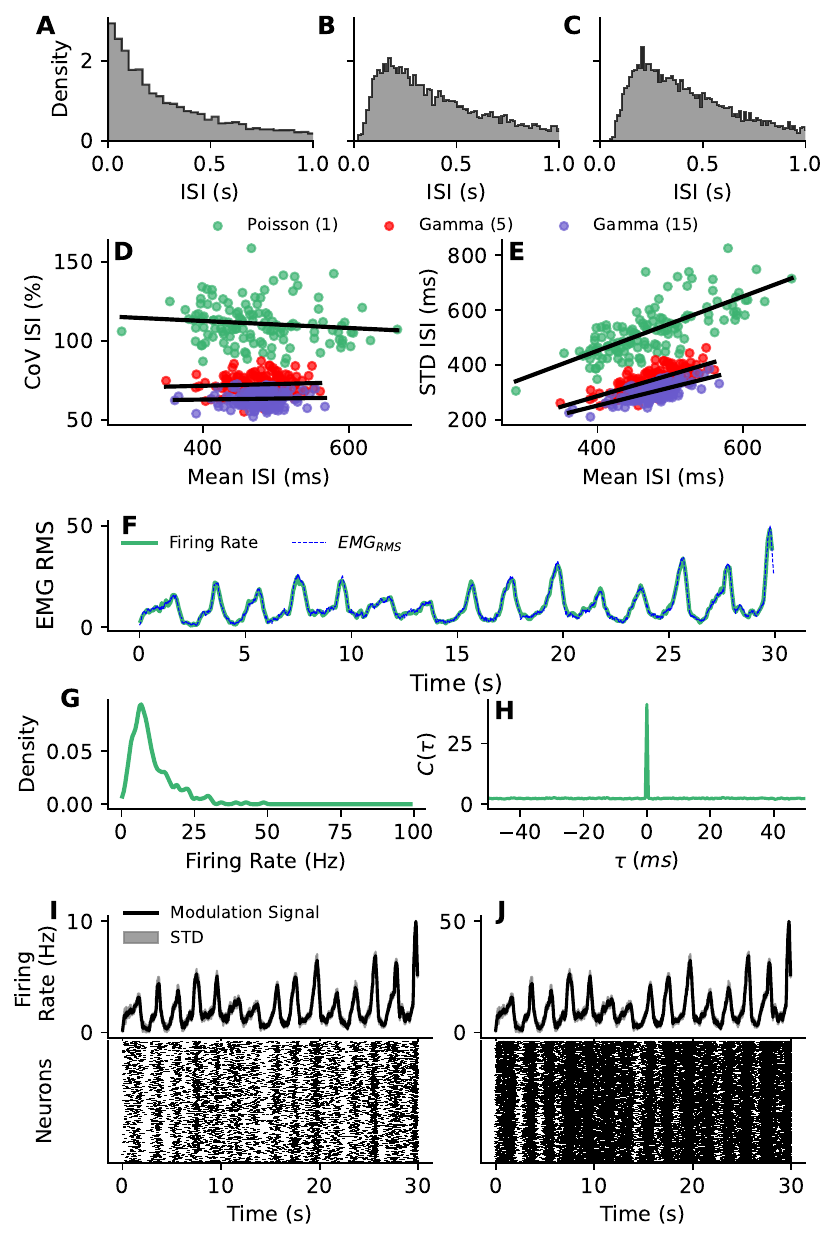}
    \caption{Spike encoding properties and associated physiological patterns. (A–C) Inter-spike interval (ISI) distributions for a Poisson process (order 1) and Gamma processes (orders 5 and 15, respectively), with an amplitude of the spike encoder's modulating signal of $\theta = 10$. (D–E) Relationship between mean ISI and variability metrics, including the coefficient of variation (CoV) and standard deviation (STD), for $\theta = 10$. (F–H) Poisson process with $\theta = 50$: (F) time-varying firing rate driven by the modulation signal, (G) firing rate distribution, and (H) autocorrelation of the active neuronal population. (I–J) Input modulation signal (black line, mean across the 128 channels; gray band, STD) and corresponding spike trains of the 128 channels (raster) for a Poisson process with $\theta = 10$ and $\theta = 50$, respectively. RMS: root mean square.}
    \label{fig:SpikeEncoding}
\end{figure}

The spike encoding patterns are shown in Figure \ref{fig:SpikeEncoding}. The figure shows an illustrative example of the encoder output for one participant in one motor task; absolute firing rates therefore depend on the normalized amplitude of that recording and are not directly comparable with the population averages reported in the main text. The Poisson process (panel A) exhibits an exponential ISI distribution with high temporal variability, whereas Gamma processes (panels B–C) increase regularity, reducing ISI dispersion as the process order increases. Accordingly, the Poisson process shows higher CoV and standard deviation, while Gamma processes progressively decrease these variability measures (panels D–E). The empirical CoV values exceed the theoretical ones ($1/\sqrt{n}$; 100, 45, and 26\%) because the input modulates the firing rate over time. These results correspond to $\theta = 10$, reflecting low firing rates.

For $\theta = 50$, the firing rate tracks the modulation signal $EMG_{RMS}$ (panel F), indicating that the encoder preserves the temporal dynamics of the input. The firing rate distribution (panel G) peaks around 7 Hz, with a broad spread that reflects the variability of the population firing rate over time. The autocorrelation function (panel H) peaks at 0 ms, consistent with the statistical independence of spikes in a Poisson process \mbox{\cite{Perkel1967}}.

Increasing $\theta$ results in higher firing rates across the 128 channels (panels I–J). For instance, at $\theta = 50$, the firing rate fluctuates between approximately 2 and 30 Hz, producing denser spike trains compared to $\theta = 10$. Overall, increasing the modulation amplitude raises firing activity while preserving the temporal structure of the input signal, demonstrating that the encoder effectively maps continuous signals into physiologically plausible spike trains.

\subsection{\texorpdfstring{Pseudo-code of the spike encoding}{Pseudo-code of the spike encoding}}
Algorithm~\ref{alg:encoder} summarizes the implementation of the spike encoder. The threshold $\xi_c$ has unit mean and follows an exponential distribution for $n = 1$, yielding a Poisson process, and a Gamma distribution (shape and rate $n$) for $n > 1$, yielding a Gamma process; therefore, the order modulates spike-train regularity without altering the mean firing rate.

\begin{algorithm}[H]
\caption{Spike encoding.}
\label{alg:encoder}
\begin{algorithmic}[1]
\REQUIRE Normalized RMS envelope $\hat{x}_c(t) \in [0,1]$ of channels $c = 1,\dots,C$; amplitude $\theta$; order $n$; sampling period $\Delta t = 1/f_s$
\ENSURE Spike trains $S_c(t) \in \{0,1\}$, $t = 1,\dots,N$
\FOR{$c = 1$ \TO $C$}
    \STATE $\Lambda_c \leftarrow 0$; \ $\xi_c \leftarrow -\frac{1}{n}\ln\prod_{k=1}^{n}U_k$, \ $U_k \sim \mathcal{U}(0,1)$
\ENDFOR
\FOR{$t = 1$ \TO $N$}
    \FOR{$c = 1$ \TO $C$}
        \STATE $\Lambda_c \leftarrow \Lambda_c + \theta\,\hat{x}_c(t)\,\Delta t$
        \IF{$\Lambda_c \geq \xi_c$}
            \STATE $S_c(t) \leftarrow 1$; \ $\Lambda_c \leftarrow 0$; \ draw a new $\xi_c$
        \ELSE
            \STATE $S_c(t) \leftarrow 0$
        \ENDIF
    \ENDFOR
\ENDFOR
\end{algorithmic}
\end{algorithm}

\section{Baseline methods configurations}
Table \ref{Tab:1} presents the reference methods implemented in this study, where the MLP and CNN architectures correspond to those described in \cite{Mendez2021}. For the RR, KNN, HGB, and SVR methods, a hyperparameter search was conducted, and the configurations with the best performance according to $R^2$ were selected.  

\begin{table}[ht]
\caption{Baseline methods implemented for comparison with the proposed method ($SNN_{Physio}$).}
\label{Tab:1}
\centering
\small
\begin{tabular}{@{}l>{\raggedright\arraybackslash}p{0.85\linewidth}@{}}
\hline
Method & Hyperparameters (search and default ranges) \\ \hline

MLP   & Hidden layer sizes: $\left\{256, 128, 64\right\}$, activation: relu, dropout: 0.2, learning rate: 0.001, weight decay: 0.001, max iterations: 100, early stopping: 13 \\

RR    & Regularization $\alpha$ $\in$ $\left\{ 0.1, 1, 10, 20, 50 \right\}$ \\

KNN   & Number of neighbors: $\left\{ 5, 15, 20 \right\}$, weights: $\left\{ \text{uniform}, \text{distance} \right\}$ \\

HGB   & Maximum iterations: $\left\{ 20, 30, 50 \right\}$, learning rate: $\left\{ 0.1, 0.05 \right\}$, maximum depth: $\left\{2, 3, 5\right\}$, max features: 0.8, early stopping: 13 \\

SVR   & Kernel: $\left\{ \text{rbf}, \text{linear} \right\}$, penalization: $\left\{ 0.1, 1, 5, 10 \right\}$, epsilon: $\left\{ 0.01, 0.1, 0.5, 1 \right\}$ \\

CNN   & \leavevmode Input: band-pass filtered sEMG (1 channel), Conv blocks: $\left\{3\times(1\rightarrow128),\ 2\times(128\rightarrow128),\ 1\times(128\rightarrow64)\right\}$, kernel: $(3,1)$, pooling: AvgPool, norm: BatchNorm, activation: ReLU, dropout: 0.5, learning rate: 0.001, weight decay: 0.001, early stopping: 13 \\ \hline

\end{tabular}\par\smallskip
\footnotesize
MLP: Multilayer Perceptron; RR: Ridge Regression; KNN: K-Nearest Neighbors; HGB: Histogram-based Gradient Boosting; SVR: Support Vector Regressor; CNN: Convolutional Neural Network.
\end{table}

\section{Hyperparameter tuning}
The results of the hyperparameter search, conducted to determine the configuration that maximizes the performance of $SNN_{Physio}$ for kinematic decoding across different conditions, are summarized in Tables \ref{Tab:hyper1FP}, \ref{Tab:hyper2FP}, and \ref{Tab:hyperFG}. Each table lists the ten highest-ranked hyperparameter combinations identified for the corresponding task (1FP, 2FP, or FG), obtained by pooling the five best-ranked distinct configurations of each participant and sorting them by $N_{\mathrm{sub}}$ and then by mean $R^2$. For each combination, the table reports the number of participants for whom it ranked among the five best configurations ($N_{\mathrm{sub}}$), its best and average ranking across participants, and the corresponding mean and maximum variance-weighted $R^2$ values. The final configuration for each task was selected among the highest-ranked combinations of the corresponding table and was subsequently applied to all participants and experimental conditions in the main analysis.

\begin{table}[ht]
\centering
\caption{Top hyperparameter combinations for the one-finger pinch task, ranked by the number of participants for whom they were among the five best configurations ($N_{\mathrm{sub}}$) and then by mean coefficient of determination ($R^2$). $\theta$: amplitude of the spike encoder's modulating signal; $V_{thr}$: baseline neuronal threshold; E/I: excitation/inhibition; WB: without E/I balance restriction.}
\label{Tab:hyper1FP}
\small
\setlength{\tabcolsep}{4pt}
\begin{tabular}{ccccccccc}
\toprule
\textbf{$\theta$} &
\textbf{$V_{thr}$} &
\textbf{Order} &
\textbf{\makecell{E/I\\Balance}} &
\textbf{$N_{\mathrm{sub}}$} &
\textbf{\makecell{Best\\ranking}} &
\textbf{\makecell{Average\\ranking}} &
\textbf{\makecell{Mean\\$R^2$}} &
\textbf{\makecell{Max\\$R^2$}} \\
\midrule
400 & 4.6 & 25 & WB & 1 & 1 & 1.00 & 0.8088 & 0.8088 \\
350 & 5.3 & 15 & WB & 1 & 2 & 2.00 & 0.8061 & 0.8061 \\
350 & 4.9 & 15 & WB & 1 & 3 & 3.00 & 0.8050 & 0.8050 \\
400 & 4.7 & 15 & WB & 1 & 4 & 4.00 & 0.8037 & 0.8037 \\
350 & 5.4 & 15 & WB & 1 & 5 & 5.00 & 0.8026 & 0.8026 \\
250 & 4.6 & 5 & 70/30 & 1 & 1 & 1.00 & 0.7984 & 0.7984 \\
250 & 4.0 & 15 & 70/30 & 1 & 2 & 2.00 & 0.7950 & 0.7950 \\
250 & 5.0 & 5 & 80/20 & 1 & 3 & 3.00 & 0.7932 & 0.7932 \\
250 & 3.6 & 5 & 70/30 & 1 & 4 & 4.00 & 0.7916 & 0.7916 \\
300 & 4.1 & 5 & 70/30 & 1 & 5 & 5.00 & 0.7908 & 0.7908 \\
\bottomrule
\end{tabular}
\end{table}

\begin{table}[ht]
\centering
\caption{Top hyperparameter combinations for the two-finger pinch task, ranked by the number of participants for whom they were among the five best configurations ($N_{\mathrm{sub}}$) and then by mean coefficient of determination ($R^2$). $\theta$: amplitude of the spike encoder's modulating signal; $V_{thr}$: baseline neuronal threshold; E/I: excitation/inhibition; WB: without E/I balance restriction.}
\label{Tab:hyper2FP}
\small
\setlength{\tabcolsep}{4pt}
\begin{tabular}{ccccccccc}
\toprule
\textbf{$\theta$} &
\textbf{$V_{thr}$} &
\textbf{Order} &
\textbf{\makecell{E/I\\Balance}} &
\textbf{$N_{\mathrm{sub}}$} &
\textbf{\makecell{Best\\ranking}} &
\textbf{\makecell{Average\\ranking}} &
\textbf{\makecell{Mean\\$R^2$}} &
\textbf{\makecell{Max\\$R^2$}} \\
\midrule
400 & 3.7 & 25 & WB & 1 & 1 & 1.00 & 0.9158 & 0.9158 \\
400 & 5.6 & 25 & WB & 1 & 2 & 2.00 & 0.9139 & 0.9139 \\
400 & 5.5 & 25 & WB & 1 & 3 & 3.00 & 0.9138 & 0.9138 \\
400 & 5.4 & 25 & WB & 1 & 4 & 4.00 & 0.9137 & 0.9137 \\
400 & 5.8 & 25 & WB & 1 & 5 & 5.00 & 0.9132 & 0.9132 \\
350 & 3.1 & 25 & 65/35 & 1 & 1 & 1.00 & 0.8895 & 0.8895 \\
350 & 3.1 & 15 & 65/35 & 1 & 2 & 2.00 & 0.8810 & 0.8810 \\
350 & 3.0 & 15 & 65/35 & 1 & 3 & 3.00 & 0.8733 & 0.8733 \\
350 & 3.5 & 15 & 65/35 & 1 & 4 & 4.00 & 0.8731 & 0.8731 \\
350 & 3.3 & 15 & 65/35 & 1 & 5 & 5.00 & 0.8712 & 0.8712 \\
\bottomrule
\end{tabular}
\end{table}

\begin{table}[ht]
\centering
\caption{Top hyperparameter combinations for the full grasp task, ranked by the number of participants for whom they were among the five best configurations ($N_{\mathrm{sub}}$) and then by mean coefficient of determination ($R^2$). $\theta$: amplitude of the spike encoder's modulating signal; $V_{thr}$: baseline neuronal threshold; E/I: excitation/inhibition; WB: without E/I balance restriction.}
\label{Tab:hyperFG}
\small
\setlength{\tabcolsep}{4pt}
\begin{tabular}{ccccccccc}
\toprule
\textbf{$\theta$} &
\textbf{$V_{thr}$} &
\textbf{Order} &
\textbf{\makecell{E/I\\Balance}} &
\textbf{$N_{\mathrm{sub}}$} &
\textbf{\makecell{Best\\ranking}} &
\textbf{\makecell{Average\\ranking}} &
\textbf{\makecell{Mean\\$R^2$}} &
\textbf{\makecell{Max\\$R^2$}} \\
\midrule
400 & 5.5 & 25 & WB & 2 & 4 & 4.50 & 0.9307 & 0.9384 \\
400 & 4.4 & 25 & WB & 1 & 1 & 1.00 & 0.9463 & 0.9463 \\
400 & 4.7 & 25 & WB & 1 & 2 & 2.00 & 0.9441 & 0.9441 \\
400 & 5.4 & 25 & WB & 1 & 3 & 3.00 & 0.9384 & 0.9384 \\
400 & 5.2 & 25 & WB & 1 & 5 & 5.00 & 0.9378 & 0.9378 \\
300 & 6.0 & 25 & WB & 1 & 1 & 1.00 & 0.9289 & 0.9289 \\
400 & 6.0 & 25 & WB & 1 & 2 & 2.00 & 0.9269 & 0.9269 \\
400 & 5.6 & 25 & WB & 1 & 3 & 3.00 & 0.9262 & 0.9262 \\
250 & 6.0 & 25 & WB & 1 & 4 & 4.00 & 0.9245 & 0.9245 \\
400 & 3.4 & 25 & 70/30 & 1 & 1 & 1.00 & 0.9196 & 0.9196 \\
\bottomrule
\end{tabular}
\end{table}

\section{Full performance comparison}
\begin{figure}[!htbp]
    \centering
    \includegraphics[width=\linewidth]{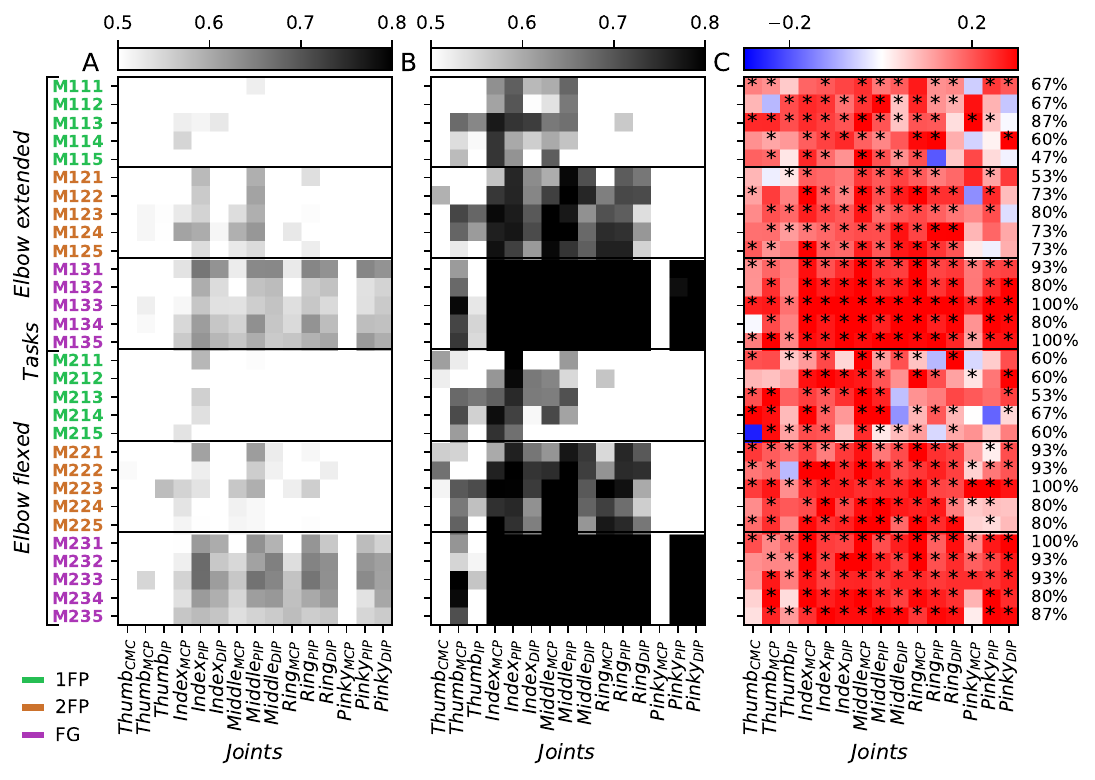}
    \caption{Simultaneous and proportional joint prediction performance across all experimental conditions. Values correspond to the median coefficient of determination ($R^2$) across subjects for each task. Rows (M$xyz$) denote the experimental conditions: elbow posture $x$ (1: extended, 2: flexed), task $y$ (1: one-finger pinch, 1FP; 2: two-finger pinch, 2FP; 3: full-hand grasp, FG), and wrist posture $z$ (1--5: from 0$\degree$ to 180$\degree$ in 45$\degree$ increments). (A) Performance of the baseline method, histogram-based gradient boosting (HGB), while (B) shows the results obtained with $SNN_{Physio}$. (C) Illustrates the difference between medians ($SNN_{Physio}$ - HGB), where red tones indicate higher performance of $SNN_{Physio}$ according to the color scale; percentages on the right indicate the proportion of joints with significant improvement in each condition. * mark statistically significant differences in favor of $SNN_{Physio}$ after false discovery rate (FDR) correction ($p < 0.05$). CMC: carpometacarpal; MCP: metacarpophalangeal; IP: interphalangeal; PIP: proximal interphalangeal; DIP: distal interphalangeal.}
    \label{fig:ComparisonMultiplesR2}
\end{figure}

The performance of simultaneous and proportional prediction across all tasks and finger joint angles is shown only for the $R^2$ metric (Figure \ref{fig:ComparisonMultiplesR2}), as it provides a representative measure of continuous kinematic decoding \cite{Ma2024, Zhang2023}. Experimental conditions are identified by a code (M$xyz$) indicating elbow posture (extended and flexed), task type (one- and two-finger pinch and full grasp), and wrist position (from 0$\degree$ to 180$\degree$). To facilitate interpretation, the figure was organized according to elbow posture and task type, while the last digit of each code represents the wrist posture: 1 for 0$\degree$, 2 for 45$\degree$, 3 for 90$\degree$, 4 for 135$\degree$, and 5 for 180$\degree$. The right side of the figure shows the percentage of finger joint angles for which the proposed method significantly outperformed HGB, with the 15 finger joint angles representing 100\% for each condition. These per-condition and per-joint comparisons used one-sided Wilcoxon signed-rank tests ($SNN_{Physio}$ $>$ HGB), with Benjamini--Hochberg false discovery rate correction across the 450 task--joint comparisons of each metric.

Overall, $SNN_{Physio}$ significantly outperformed the best baseline method (HGB) in 78\% of the 450 task–joint combinations and achieved a higher median $R^2$ in 96\% of them, with statistically significant differences favoring $SNN_{Physio}$ in all experimental conditions (see Panel C). Among the evaluated tasks, FG exhibited the highest proportion of significantly improved joints (91\% overall), reaching values between 80\% and 100\% across wrist postures. In contrast, 1FP showed the lowest percentage of significantly improved joints (63\% overall), ranging from 47\% to 87\%, whereas 2FP presented an intermediate behavior (80\% overall), with improvements varying substantially across wrist postures (approximately 53–100\%). At the joint level, significant improvements were predominantly observed for the index, middle, and ring fingers, particularly during the FG task, where their MCP, PIP, and DIP joints improved in all 10 FG conditions. In contrast, the thumb CMC and IP joints exhibited fewer significant differences (15 of the 30 conditions each), while the pinky joints showed improvements primarily during the FG task. These findings indicate that $SNN_{Physio}$ provides its greatest advantage during coordinated multi-finger movements, while simpler pinch tasks exhibit a more heterogeneous improvement pattern.

Regarding limb posture, no consistent effect of elbow configuration was observed, as both the extended and flexed elbow conditions exhibited comparable proportions of significantly improved joints (76\% and 80\%, respectively). Similarly, wrist posture had a limited influence on the relative performance of $SNN_{Physio}$, with proportions of significantly improved joints ranging from 73\% (135$\degree$) to 86\% (90$\degree$).

Overall, these findings suggest that $SNN_{Physio}$ provides greater benefits for the central fingers and coordinated multi-finger grasping tasks, with its strongest relative advantage occurring under the pronated wrist posture. These observations are based on the overall trends shown in the figure; however, more detailed analyses are required to further characterize the influence of individual wrist and elbow postures on decoding performance.

\section{\texorpdfstring{Complementary neural analysis across tasks}{Complementary neural analysis across tasks}}
To complement the neuronal characterization of the main text, three population measures were computed for each layer of $SNN_{Physio}$ from the spike trains of the test set (approximately 3 s), using the same conditions (extended elbow and 90$^{\circ}$ wrist posture) and one value per participant (averaged across seeds). The sCST was obtained by summing the spike trains of all neurons in a layer, normalizing by the number of neurons, and smoothing with a 25 ms Hann window. Population synchrony was quantified with the Golomb measure ($\chi^2$), defined as the variance of the population-averaged spike train divided by the mean variance of the individual spike trains, which ranges from 0 (asynchronous) to 1 (fully synchronous) \cite{Golomb2007}. The common input to each layer was estimated as the magnitude-squared coherence, averaged over 0–5 Hz, between the composite spike trains of two randomly selected, non-overlapping groups of neurons of equal size (sCST coherence) \cite{Negro2011}. This coherence was computed for pool sizes (number of neurons per group) ranging from 2 to half of the layer, and its spectrum is shown for the largest pool size. Finally, the gain was defined as the ratio between the mean sCST of layer 2 and that of layer 1, and quantifies the change in the population activity from the first to the second hidden layer. Differences across tasks were assessed for each layer with the statistical procedure described in the main text.

\begin{figure}[!t]
    \centering
    \includegraphics[width=\linewidth]{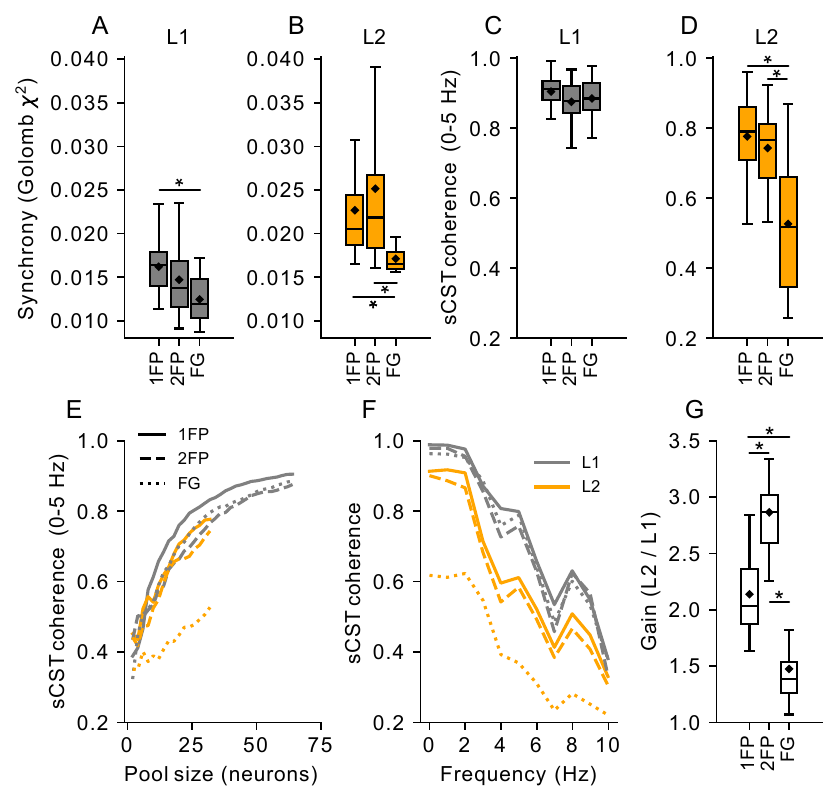}
    \caption{Population measures of $SNN_{Physio}$ by task (one-finger pinch, 1FP; two-finger pinch, 2FP; full-hand grasp, FG) in the conditions with extended elbow and 90$^{\circ}$ wrist posture. (A–B) Golomb synchrony ($\chi^2$) for layer 1 and layer 2. (C–D) Smoothed composite spike train (sCST) coherence (0–5 Hz) at the largest pool size for layer 1 and layer 2. (E) sCST coherence as a function of the number of pooled neurons (layer 1, gray; layer 2, orange; tasks by line style). (F) Coherence spectrum at the largest pool size. (G) Gain between layers (layer 2/layer 1). Boxplots summarize one value per participant (n = 23). * denotes statistically significant differences ($p<$0.05).}
    \label{fig:NeuralComple}
\end{figure}

All measures differed across tasks, except the sCST coherence of layer 1 (Figure \ref{fig:NeuralComple}). Synchrony was low in both layers (median $\chi^2\leq$ 0.022), indicating predominantly asynchronous firing, but differed across tasks in layer 1 ($p$ = 0.005, $W$ = 0.23) and layer 2 ($p<$ 0.0001, $W$ = 0.43). In layer 2, 1FP (0.021 [0.019, 0.024]) and 2FP (0.022 [0.018, 0.027]) showed higher synchrony than FG (0.016 [0.016, 0.018]) ($p<$ 0.001, $r_{rb}\geq$ 0.83), whereas in layer 1, 1FP showed slightly higher synchrony than FG ($p$ = 0.001, $r_{rb}$ = 0.79). The sCST coherence did not differ across tasks in layer 1 ($p$ = 0.20, $W$ = 0.07), but did in layer 2 ($p<$ 0.001, $W$ = 0.32), where 1FP (0.79 [0.71, 0.86]) and 2FP (0.77 [0.66, 0.81]) showed higher coherence than FG (0.52 [0.35, 0.66]) ($p<$ 0.001, $r_{rb}\geq$ 0.81). In all tasks and layers, the coherence increased with the number of pooled neurons and was concentrated below 3 Hz. Finally, the gain differed across tasks ($p<$ 0.0001, $W$ = 0.92), being higher for 2FP (2.87 [2.60, 3.02]) than for 1FP (2.03 [1.87, 2.36]) and FG (1.38 [1.26, 1.53]) ($p<$ 0.0001, $r_{rb}\geq$ 0.96), and higher for 1FP than for FG ($p<$ 0.0001, $r_{rb}$ = 1.00).

These results indicate that, in addition to firing rates and ISI statistics, the population organization of the SNN depends on the decoded task. The increase of coherence with pool size reproduces the signature of common synaptic input observed in motoneuron pools \cite{Negro2011, Farina2015}, which in the SNN arises because all neurons receive the same encoded sEMG channels. In layer 2, the higher synchrony and coherence of 1FP and 2FP compared with FG agree with the higher layer 2 firing rates reported for these tasks in the main text, and the gain between layers followed the same order as these firing rates (2FP $>$ 1FP $>$ FG). Finally, because the coherence was estimated from short segments of activity (about 3 s), the differences between tasks are more informative than the absolute coherence values.

\end{document}